\documentclass{aa} 
\bibpunct{(}{)}{;}{a}{}{,} 
\usepackage{graphicx}
\usepackage{txfonts}
\usepackage{lscape}
\usepackage{natbib,twoopt}
 \bibpunct{(}{)}{;}{a}{}{,} 
 \newcommandtwoopt{\citeads}[3][][]{\href{http://adsabs.harvard.edu/abs/#3}{\citealp[#1][#2]{#3}}}
 \newcommandtwoopt{\citepads}[3][][]{\href{http://adsabs.harvard.edu/abs/#3}{\citep[#1][#2]{#3}}}
 \newcommandtwoopt{\citetads}[3][][]{\href{http://adsabs.harvard.edu/abs/#3}{\citet[#1][#2]{#3}}}
 \newcommandtwoopt{\citeyearads}[3][][]{\href{http://adsabs.harvard.edu/abs/#3}{\citeyear[#1][#2]{#3}}}

\usepackage[ 
  breaklinks=true,
  colorlinks=true,         
   urlcolor=pdfurlcolor,    
   filecolor=pdffilecolor,  
   linkcolor=pdflinkcolor,  
   citecolor=pdfcitecolor,  %
]{hyperref} 
  
\begin{document}

\definecolor{pdfurlcolor}{rgb}{0,0,0.6}
\definecolor{pdffilecolor}{rgb}{0.7,0,0}
\definecolor{pdflinkcolor}{rgb}{0,0,0.6}
\definecolor{pdfcitecolor}{rgb}{0,0,0.6}

\newcommand{\xmm}{{\it XMM-Newton}}
\newcommand{\erosita}{{eROSITA}}
\newcommand{\rosat}{{ROSAT}}
\newcommand{\einstein}{{\it EINSTEIN}}
\newcommand{\chandra}{{\it Chandra}}
\newcommand{\swift}{{\it Swift}}
\newcommand{\gaia}{{\it Gaia}}
\newcommand{\etal}{et al.}
\newcommand{\nh}{\mbox {$N_{\rm H}$}}

\newcommand{\ms}[1]{\textcolor{cyan} {MS: #1}}

  \title{\erosita\,(eRASS1) study of the Canis Major overdensity: Developing a multi-wavelength algorithm for classifying faint X-ray sources
}
   \subtitle{}
    \titlerunning{\erosita\,(eRASS1) Study of CMa~Overdensity}
   \authorrunning{S.~Saeedi et al}
   \author{Sara Saeedi\inst{1},
           Manami Sasaki\inst{1}, Jonathan Knies\inst{1}, Jan Robrade \inst{2}, Theresa Heindl\inst{1}, Aafia Zainab\inst{1}, Steven H\"ammerich\inst{1}, Martin Reh\inst{1}, Joern Wilms\inst{1} }

\institute{\\\inst{1}  Dr. Karl Remeis-Sternwarte, Erlangen Centre for Astroparticle Physics, Friedrich-Alexander-Universit\"at Erlangen-N\"urnberg, Sternwartstrasse 7, 96049, Bamberg, Germany\\
    \email{sara.saeedi@fau.de}\\
    \inst{2}  HamburgerSternwarte, Gojenbergsweg112, 21029, Hamburg, Germany}
 

  \date{Received DATE; accepted DATE}

\abstract{}{}{}{}{} 
 
  \abstract
   {Using data from \erosita\,(extended Roentgen Survey with an Imaging Telescope Array) on board Spektrum-Roentgen-Gamma (Spektr-RG, SRG) taken during the first \erosita\, all-sky survey (eRASS1), we performed the first X-ray classification and population study in the field of the Canis~Major overdensity\,(CMa~OD), which is an elliptical-shaped stellar overdensity located  at $l= -240^{\circ}, b=-80^{\circ}$. }
   {This study aims to identify the X-ray sources in CMa~OD. We developed a classification algorithm using  multi-wavelength criteria as a preliminary method for the classification of faint X-ray sources, specifically in  regions with a high source number density.}
   {We used the brightness of the multi-wavelength counterparts (mainly from infrared and optical catalogues), along with the X-ray flux and X-ray  hardness ratios (HRs) to classify  the sources.}
   {Out of a total number of 8311 X-ray sources, we classified 1029 sources as  Galactic stars and  binaries in the foreground, 946 sources as  active galactic nuclei (AGNs) and galaxies in the background, and 435 sources with stellar counterparts that may belong to either the MW or CMa~OD. Among the sources with a stellar counterpart, we identified 34 symbiotic star candidates, plus 335 sources, of which the infrared (IR) counterparts have properties of M-giants in CMa~OD. 
   Moreover, there is a known high-mass X-ray binary (HMXB, 4U\,0728-25) in the field of our study;  according to the \gaia\,parallax of its companion, it appears to be a member of CMa~OD. There is also a recently detected transient low-mass X-ray binary\,(LMXB, SRGt\,J071522.1-191609) is also present; it may be a member of CMa~OD based on its companion, which is most likely highly absorbed and is thus located behind the Galactic disk. In addition, we present the X-ray luminosity function\,(XLF) of members and candidate members of  CMa~OD. It is dominated by sources with luminosities of $<2\times10^{32}$--$10^{33}$~erg\,s$^{-1}$ in the energy range of 0.2--2.3~keV. These sources are expected to be either accreting white dwarfs or quiescent LMXBs.} 
   {}


   \maketitle
%

\section{Introduction}
\label{intro}
Today, we know of almost 60 satellite galaxies of the Milky Way (MW) and $\sim$40 satellite galaxies of M31
\citep{2012AJ....144....4M}.
With the most prominent exception of the Large and Small Magellanic Clouds, 
which are dwarf irregular galaxies with copious star formation, most of
these systems are metal-poor dwarf spheroidal galaxies (dSphs)  with
metallicities as low as $[\mathrm{Fe}/\mathrm{H}]<-3$ \citep[e.g.][]{2010Natur.464...72F,2012AJ....144....4M}. This population
of extremely metal-poor stars makes the dSphs ideal laboratories
for the study of the stellar populations in the earliest epochs of chemical
enrichment in the universe. Unlike other types of nearby galaxies, the X-ray population of these old, low-mass satellite galaxies of the MW has been poorly studied. We expect that the main population of late-type stars of dSphs have formed white
dwarfs \citep[e.g.][]{2009ARA&A..47..371T}. Therefore, we would expect a mixed population of accreting with dwarfs\,(AWDs) and transient low-mass X-ray binaries (LMXBs) with $L<10^{35}\,\mathrm{erg}\,\mathrm{s}^{-1}$, which is very important for two reasons. First, the close distance of nearby dSphs provides a unique opportunity to study the X-ray luminosity function (XLF) of AWDs, which is not well understood in the case of Galactic sources due to the varying distances and selection effects. Second, it will improve our understanding of transients, which are expected to have luminosities between X-ray binaries (XRBs) and AWDs \citep{2006A&A...450..117S}. On the other hand, the population of low-luminous X-ray sources is a mix of cataclysmic variables\,(CVs), symbiotic stars,  quiescent LMXBs, and magnetically active binaries\,(ABs) More details are provided in the literature\citep[e.g.  ][]{2022A&A...661A..35S}. A deep multi-wavelength study of these X-ray sources in a nearby dSph will provide us with a pristine sample of each type of the low-luminosity X-ray sources and allow us to constrain the contribution of each type to the total low-luminosity population and thus to better understand the XLF.

We started a study of the X-ray source population of dSphs around the
Milky Way. We measured the XLF in the hard and soft X-ray bands for Draco dSph, as one of the oldest nearby dSphs with an age of $\sim 10^{10}$ yr \citep[e.g.][]{2002AJ....124.3222B}.
The XLF of sources in the soft energy range (0.2--2.0\,keV) shows an excess above the AGN
distribution for low luminosity X-ray sources ($<10^{34}\,\mathrm{erg}\,\mathrm{s}^{-1}$).
Our additional, more detailed classification of the X-ray population of
 Draco dSph has shown that the main X-ray sources in the Draco dSph
are AWDs \citep{2018MNRAS.473..440S,2019A&A...627A.128S}. Following up on the significant detection of AWDs in the Draco dSph,
we have applied a similar analysis method to other dSphs with
different stellar masses and star formation histories to verify the
differences of the population of AWDs in these galaxies. We studied
the X-ray sources of the Willman\,1 dSph, one of the least massive
dSphs, using three archival \xmm\, observations, confirming the
presence of a $\beta$-type symbiotic binary and some candidates of
transient LMXBs \citep{2020MNRAS.499.3111S}. We also confirmed the presence of multiple AWDs and transient LMXBs in Sculptor dSph, which is more massive than Draco,\citep{2022MNRAS.512.5481S}. Now, the \erosita\, all-sky survey (eRASS) gives us a unique opportunity to study the X-ray population of the satellite galaxies of the Milky Way.

CMa~OD is a stellar population, which has been suggested to be the main core of a dSph. It is located very close to the Galactic plane RA=$07\fh~12\fm~35.00\fs$, Dec=$-27^{\circ} 40\arcmin 00\arcsec$ [J2000]) \citep{2012AJ....144....4M}, and at a distance of 7.0$\pm$1.0~kpc from the Sun. This distance measurement has been also confirmed by the later study of \citep{2021MNRAS.501.1690C}. In fact, the origin of the CMa~OD is unclear. One hypothesis is that this overdesity is the main part of the core of the most massive and the closet dwarf galaxy to the MW. This hypothesis is supported by a certain degree of evidence; for instance, 
the old stellar population of M-giant in this overdesity is similar to that of other nearby dSphs \citep{2004MNRAS.348...12M}. In addition, the presence of some globular clusters around the CMa~OD reinforces the idea of a possible accreted
galaxy within the inner Galactic halo \citep[e.g..][]{2010MNRAS.404.1203F} as well as the excess of bright blue stars in the central regions of CMa, with respect to the Galactic plane, shows the extra-Galactic origin of CMa~OD \citep[e.g.][]{2007ApJ...662..259D}. The subtracted-density map, where the Galactic stars have been filtered out,
indicates a dense M-giant population of $\sim$4500 stars kpc$^{-2}$ in the core of the CMa~dSph \citep{2006MNRAS.366..865B}. Deep optical and infrared (IR) population studies show that the luminosity, dimensional properties, and the population of the red giant branch of CMa~dSph is very similar to that of Sagittarius dSph, in a way that Sagittarius dSph would look like CMa~dSph if it was placed at the distance of CMa~OD.  On the other hand, there are studies on the stellar population, metallicity of this overdensity, and special distribution of blue stars, which weaken the above evidence and show that the CMa~OD can be a part of the Galactic wrap \citep[see more details in][]{2018MNRAS.473..647D, 2006A&A...451..515M,2021MNRAS.501.1690C}. Therefore, it seems that no ultimate conclusion has yet been confirmed for CMa~OD.

In general, the classification of low-luminosity X-ray sources with large positional uncertainties is  a complicated task. Hence, the classification of X-ray sources in the field of CMa~OD, which is located behind the Galactic disk with a very high population density of Galactic stars and binary systems in the foreground is challenging. Therefore, we decided to develop an algorithm to carry out a systematic analysis and thus increase the probability of the classification of faint X-ray sources using the multi-wavelength data. Recently, two different approaches of big data analysis and machine learning have tried to improve the classification of X-ray sources. First, \citet{2022ApJ...941..104Y} have provided a machine learning method based on the properties of known X-ray sources to classify the unknown sources in the  \chandra\, Source Catalogue. It provides the first machine learning approach to analyse the different properties of the different classes of X-ray sources. Other studies have presented more general approaches for source classification \citep[e.g.][]{2018MNRAS.473.4937S, 2017PASP..129f2001M, 2011A&A...534A..55S}. \citet{2022ApJ...941..104Y}  selected optical and IR counterparts of the X-ray sources within a circle with a radius of 10" as the positional uncertainty and assumed the nearest optical or IR source as the counterpart(s) or the X-ray source. This is not as reliable as the method from \citet{2018MNRAS.473.4937S}, a code known as NWAY that was developed to determine  the most probable counterpart of an X-ray source from cross-matching multiple catalogues. For the source classification, we have used the NWAY method searching for IR and optical counterparts supporting with multi-wavelength algorithm as it is explained in Sects.\,\ref{nway} and\,\ref{algorithm-sec}.

In this paper, we first explain the characteristics of sources of the first \erosita\,\citep{2021A&A...647A...1P} all-sky survey (eRASS1) catalogue \citep{2024A&A...682A..34M} in the field of CMa~OD (Sect.\,\ref{data-sec}). Next, we present the details of the multi-wavelength catalogues, which have been used in this study (Sect.\,\ref{multi-sec}). Section\,\ref{algorithm-sec} explains  the algorithm of the multi-wavelength  classification, which has been developed in this work. Section\,\ref{xray-sect} discusses the X-ray properties used for the classification. Section\,\ref{diss-sec} presents the results of the classification and the XLF of the low-luminosity X-ray sources of CMa~OD.

\section{Data selection and methodology}
\label{data-sec}
\subsection{eRASS1 data of CMa}
\begin{figure}
\includegraphics[trim={0.8cm 2.cm 2.5cm 0.cm},clip,width=0.5\textwidth]{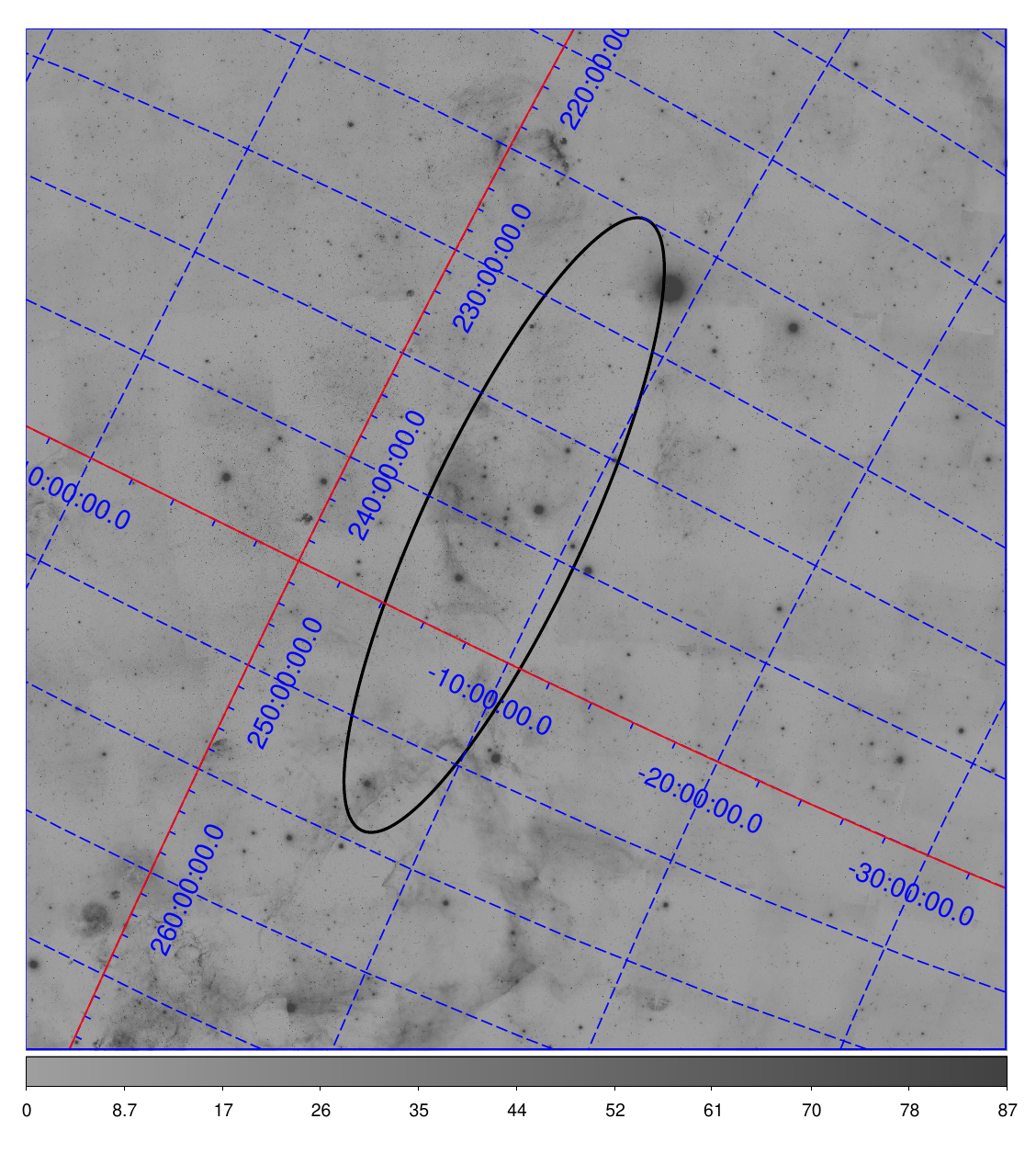}
\caption{Optical image of the red band of DSS~II survey \citep{2000ASPC..216..145M} of the field of CMa~OD. Center of the CMa~OD
  (black ellipse). The ellipse covers
  central core of the CMa~dSph according to \citet[][]{2004MNRAS.348...12M} and has a semi-major axis of
  ${\sim}20^{\circ}$. }
  \label{cma-image}
\end{figure}
In our study of CMa~OD, we considered all detected eRASS1 sources located in the field of  CMa~OD (${\sim}20$\,deg$^{2}$). Figure\,\ref{cma-image} shows the region of the core of CMa~OD. This area  has been determined as the core of CMa~dSph in the IR study of \citet{2004MNRAS.348...12M} and the same area has been applied in the work of \citet{2021MNRAS.501.1690C}. We have a total of 8311 eRASS1 sources.
The sources in the field of CMa~OD were directly taken from the main catalogue of eRASS1, which includes sources with a detection maximum likelihood of ML $\geq$6\,(>3$\sigma$) \footnote{The detection likelihood ($L$) is calculated from Poissonian probability\,($L=-ln(probability)$), based on aperture count extraction. This probability is calculated using the raw counts of the source and the raw counts of the background maps taken from  an aperture representing $75\%$ of the encircled-energy fraction radius of the point speared function \citep[see][]{2024A&A...682A..34M, 2022A&A...661A...1B}.}. According to \citet{2024A&A...682A..34M}, $\sim$14$\%$ of these sources are potentially spurious, while the probability of spurious sources decreases to $\sim$1$\%$ if the criterion is changed to ML $\geq10$\,(>4$\sigma$). In the first step, we have started our work selecting all catalogue sources  with ML $\geq$6. More than half of these sources are in the range of 6 $<$ML$<$ 10, 
the majority being AGNs or Galactic stars, which are useful for classification based on multi-wavelength counterparts (as explained in Sect.\,\ref{algorithm-sec}). 
For those sources, which our analysis has identified  as potential members of CMa~OD (see Sect.\,\ref{algorithm-sec}), we applied the more strict criterion of  ML$\geq$10\,(>4$\sigma$). Otherwise, these sources have been flagged as candidate members of CMa~OD. Using a similar approach to the way \citet{2024A&A...682A..34M} performed a systematic astrometry correction based on quasars in the WISE catalogue, we used the eRASS1 catalogue coordinates to search for counterparts.
\subsection{Probabilistic cross-matching method}
\label{nway}
We have used NWAY to search for the optical and IR counterparts
of the X-ray sources (as explained in Sect.\,\ref{algorithm-sec}). It is based on  Bayesian statistics and computes the matching probability by taking into account such  parameters as the positional uncertainties of the X-ray source and its counterpart, the distance of a counterpart from the X-ray source, and  the number densities of available counterparts \citep{2018MNRAS.473.4937S}. The NWAY code provides a ({\selectfont p-i}) parameter for each match \citep[][]{2018MNRAS.473.4937S}, which gives the probability to be the likely counterpart of a source. Using the NWAY, \citet{2022A&A...661A...3S} performed a multi-wavelength study to classify the point-like X-ray sources detected in the \erosita\, Final Equatorial-Depth Survey. Their main aim was to establish criteria for the classification of background sources and to present a clean sample of active galactic nuclei (AGNs). Therefore, they have excluded the sources, which are near the Galactic plane (|$b$|~<~15~deg) and only have taken to account the sources with  optical or IR counterparts with a high NWAY  match probability ({\selectfont p-i}>90$\%$). However,  using only the parameter of match probability to select the counterpart is not practical in a crowded field close to the Galactic plane. Therefore, we have developed an algorithm based on the multi-wavelength investigation of probable counterparts, which is tailored for the classification of (particularly) faint X-ray sources in a dense and crowded field such as the field of CMa~OD (See Sect.\,\ref{algorithm-sec}). 
 \subsection{Problem of optical loading}
\begin{figure}
\includegraphics[trim={0.cm 0.0cm 0.0cm 0.cm},clip,width=0.5\textwidth]{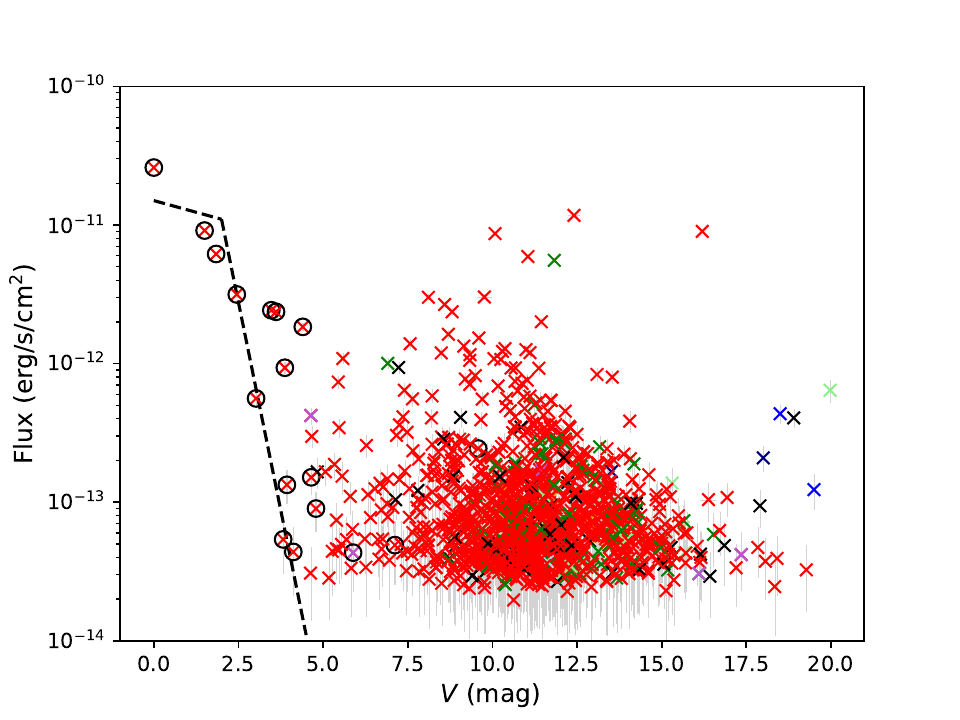}
\caption{Flux\,($0.2-2.3$~keV) of the X-ray sources versus the optical magnitude\,($V$) of their counterparts from the SIMBAD data base. Galactic sources are labeled with a red cross. The AGNs and quasars are shown with blue crosses. Stellar objects for which the distances are not precisely estimated are shown with green crosses. The dashed line presents the X-ray flux threshold, which is caused by optical loading \citep[]{2024A&A...682A..34M}. The sources with the optical loading problem are labelled with an extra black ring and  their list is presented in Table \ref{optload}. \label{optical-loading}}
\end{figure}

It is known that bright objects in the optical and IR can cause spurious signals in X-ray observations, where optical and IR photons can trigger X-ray CCDs and wrongly recorded as X-ray photon events; this is is referred to as the optical loading problem. Therefore, we have filtered out optical and IR bright objects, which falsely show very soft X-ray emission, following the method from \citet[][]{2024A&A...682A..34M}, as shown in Fig.\,\ref{optical-loading}. We have a total of 17 sources in our full sample 
and all of them are bright stars  
in SIMBAD Astronomical Database \citep[hereafter, SIMBAD,] []{2000A&AS..143....9W}.  These sources, which are potentially affected by optional loading are mainly identified based on their $V$ mag value of $V\le4.5$~mag. Five additional ones were flagged due to their IR ($J$) magnitude and in one case this was due to the blue ($B$) magnitude.
These sources are excluded in our study, however they were kept in the catalogue and have been flagged with "opt.loading" label in the comment column of Table \ref{final-cata}. The list of the excluded sources with the ID of our catalogue and their astronomical names is presented in Table.\,\ref{optload}. The $V$ magnitude of the sources is provided. If the $J$ or $B$ magnitudes are caused by the optical loading, these magnitudes have also been presented. 
\begin{table}
 \centering
\hspace{-0.5cm} \caption{$V$ magnitude of optical-loading sources\label{optload}}
     \begin{tabular}{lll}
\hline\hline
Source No.& Astronomical name&  Optical magnitude(s)$^{*}$\\
\hline
178& alf\,CMa,&$V$= -1.46\\
917&omi\,01\,CMa& $V$=3.78\\ 
1147&mu.\,CMa\,A&$V$=5.09, $J$=2.64\\ 
1488&eps\,CMa& $V$=1.50\\
1915&sig\,CMa& $V$=3.47\\
2095&omi02\,CMa& $V$=3.02\\
2197&gam\,CMa& $V$=4.12\\
2871&del\,CMa& $V$=1.84\\
3809&27\,CMa&$V$=4.65, $B$=4.45\\
3893&ome\,CMa& $V$=3.82\\
3894&HD\,56618& $V$=4.64, $J$=1.30\\
4171&145\,CMa& $V$=4.79, $J$=1.91\\
4527&tau\,CMa& $V$=4.40\\
4939&V~MZ\,CMa& $V$=5.88, $J$=2.26\\
5417&eta\,CMa& $V$=2.45\\
7642&l\,Pup&$V$=3.93\\
7781&c\,Pup& $V$=3.61\\    
      \hline
     \end{tabular}
\raggedright{*: The optical and IR magnitudes of the sources are taken from \citet{2002yCat.2237....0D}. The source No. is the ID of the sources in the catalogue of this work, see Sect.\,\ref{catalogue-x-ray}}. 
\end{table}

A sample of the X-ray source catalogue of this work is shown in Appendix\,\ref{final-cata}, while the full catalogue is available online. Our catalogue presents the information of each source including eRASS1 ID, right ascension\,(RA), declination,(Dec), positional uncertainty ($r$), the source rate and its corresponding flux in 0.2--2.3~keV, total detection ML in the same energy range, hardness ratios\,(HR) (see sect.\,\ref{hr-sec}), the class of the sources based on this study, and a comment column; these comments  mainly give the astronomical name of  a source if it is an already known source according to SIMBAD or some additional necessary information related to (e.g. optical loading problem, pile-up, etc). Our source list is sorted based on RA and a number\,(No.)\,
is given to each source for presentation in this work.

\section{Multi-wavelength catalogues}
\label{multi-sec}
For the multi-wavelength studies, we considered the most recent available all-sky catalogues. The X-ray sources have been checked via SIMBAD to identify the known sources in the field of the study. Also, the most updated AGN or galaxy catalogues helped to identity the known background counterparts. No matter whether a source has been already classified or not, the optical and IR properties of maximum three counterparts of the X-ray source were collected for  further analysis (see Sect.\,\ref{algorithm-sec}). The details of the catalogues used in this work are presented below.

\begin{figure*}
\includegraphics[trim={0.cm 0.0cm 0.0cm 0.cm},clip,width=0.5\textwidth]{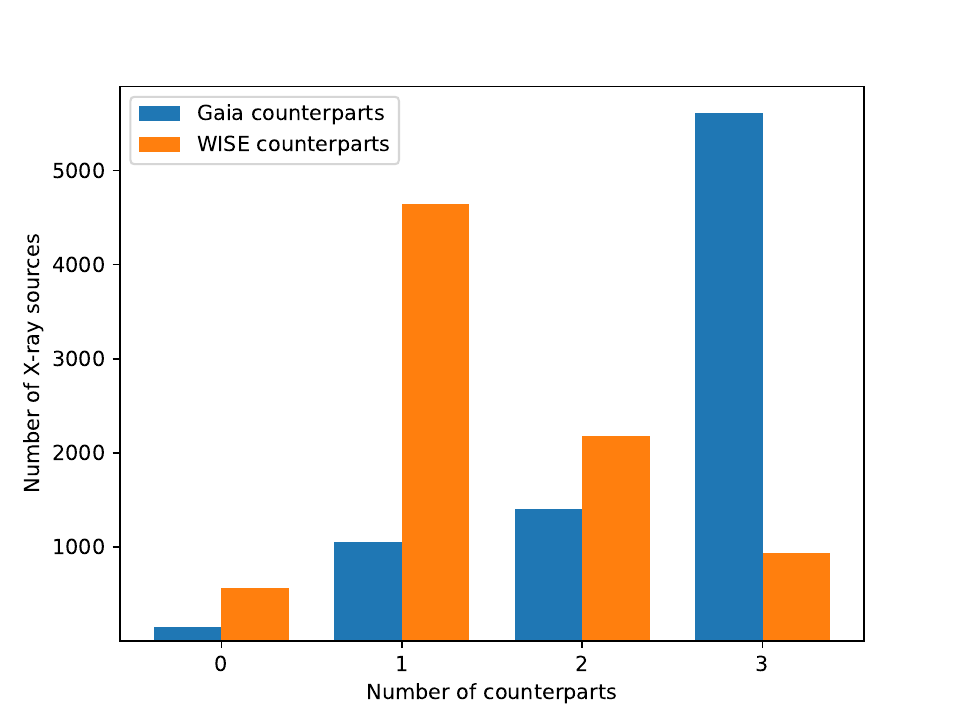}
\includegraphics[trim={0.cm 0.0cm 0.0cm 0.cm},clip,width=0.5\textwidth]{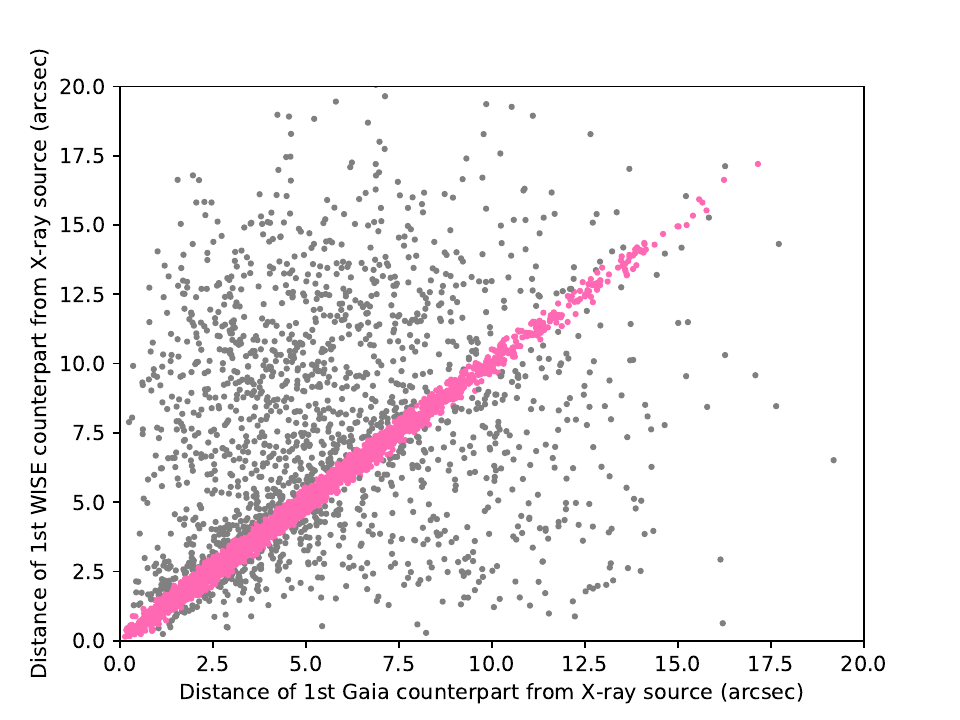}
\caption{Number of X-ray sources in the field of CMa~OD over the number of their counterparts (left). The plot shows the number of optical \gaia\,counterparts (in blue) and IR WISE counterparts (in orange) within the $3\sigma$ positional error of the X-ray sources. The plot on the right shows the distance of the first \gaia\,and WISE counterparts from the X-ray source in the field of CMa~OD. Pink sources have first \gaia\,counterpart within the positional uncertainty of the first WISE counterparts, which can thus be considered to be the same source.}\label{num-counter}
\end{figure*}
\subsection{IR counterparts of the sources}
\label{infra-sec}
We searched for mid-infrared (MIR) counterparts in the WISE all-sky survey catalogue in four energy bands \citep[3.4, 4.6, 12, and 22 $\mathrm \mu$m, known as $W1$, $W2$, $W3$, and $W4$, respectively;][]{2014yCat.2328....0C}.  The catalogue of \citet{2014yCat.2328....0C} includes the near-infrared (NIR) counterpart of the 2MASS All-Sky Survey Catalogue in the three $J$, $H$, $K$ standard bands \citep{2003yCat.2246....0C}. In the direction of CMa~OD, we applied the average extinction of 0.28, 0.17, and 0.12 for the $J$, $H$, $\text{and }K$ bands, respectively \citep{2011ApJ...737..103S}.  The extinction for the IR WISE bands was negligible \citep{2011ApJ...737..103S}. The IR counterparts are crucial for the source classification as discussed in Sect.\,\ref{algorithm-sec}.
\subsection{Optical counterparts of the sources}
\label{optical-sec}
The main catalogue of our optical investigation is from third \gaia\,data release \citep[DR3,][]{2023A&A...674A...1G}. The advantage of \gaia\,is that it has parallax measurements which can be used for identifying Galactic foreground stars. 
Source distances based on the \gaia\,parallax have been taken from \citet{2021AJ....161..147B}, who applied a probabilistic approach using a three-dimensional model for the Milky Way to measure the distance. The model considers the interstellar extinction and \gaia's  magnitude limit, which varies
from 19.2 mag in the Galactic center to 20.7 mag for most of the rest of the sky. 
\citet{2023A&A...674A...1G} reported the magnitudes of the sources in three filters of $G$ mag (roughly $\lambda$=300--1050\,nm), $G_{BP}$\,($\lambda$=330--680\,nm) mag, and  $G_{RP}$\,($\lambda$=640--1050\,nm), which have been used in our study. 
To identify  Galactic binary systems, we use the results of \citet{2021MNRAS.506.2269E}, who has provided a catalogue of a million Galactic binary systems within a distance of $\sim$1kpc.

\begin{figure*}[!ht]
\includegraphics[trim={4.cm 7.0cm 3.cm 0.cm},clip,width=1.\textwidth]{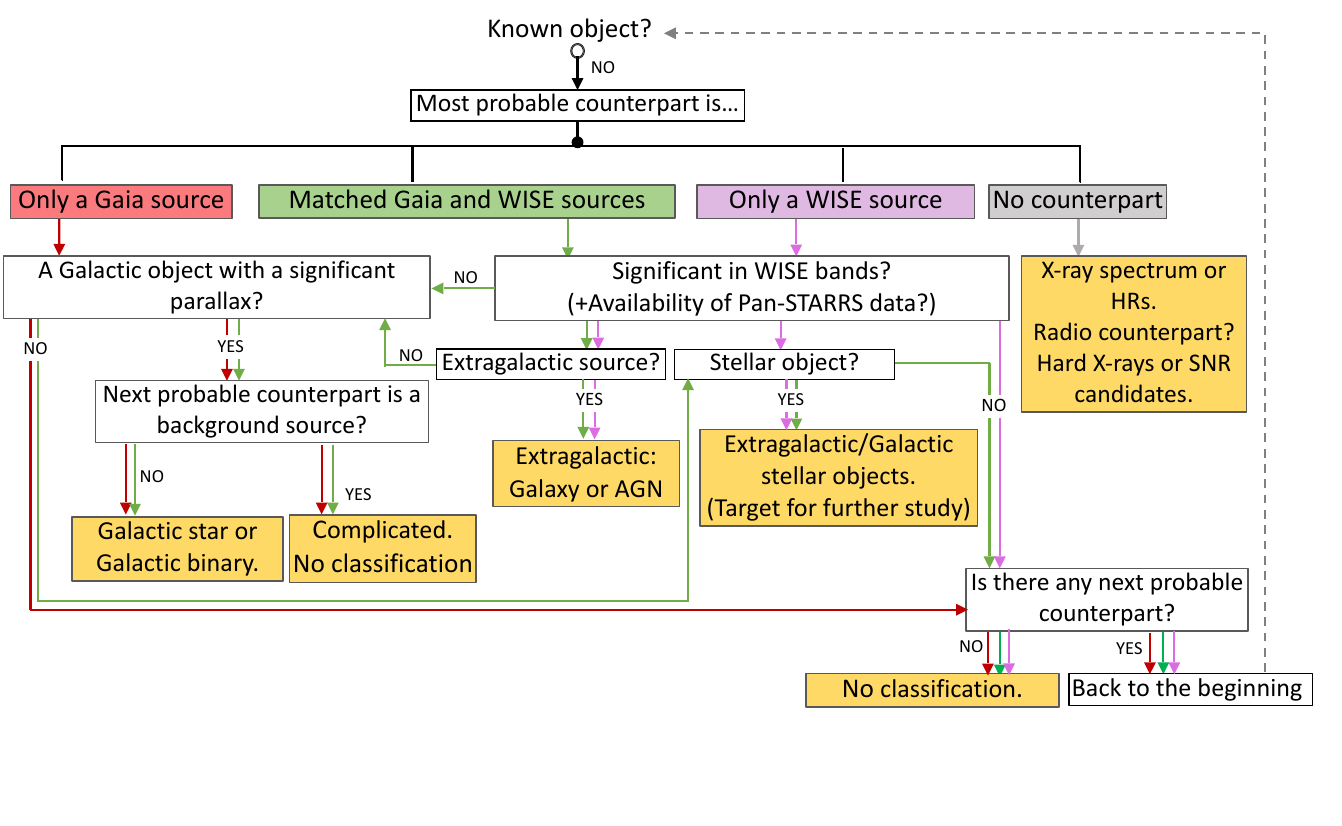}
\caption{Algorithm for the classification of X-ray sources based on multi-wavelength information.  Each X-ray source is run through the pipeline based on the conditions of their optical and IR counterparts. Details are explained in Sect.\,\ref{algorithm-sec}.  \label{algorithm}}
\end{figure*}

Moreover, the data of Pan-STARRS1 survey have been used to check the characteristics of background sources. The survey presents the optical data in 5 bands of $grizy$\,(495.7, 621.1, 752.2, 867.1, and 970.7\,nm, respectively).
\citet{2022A&A...661A...3S}  used The Kilo-Degree Survey\,(KiDs) \citep{2019A&A...625A...2K}, DESI Legacy Imaging Surveys\,(LS8) \citep{2019AJ....157..168D}, and Hyper Suprime-Cam\,(HSC) \citep{2018PASJ...70S...8A} data 
for the  optical studies of background sources. None of these surveys cover the field of CMa~OD. However, the similar sensitivity of $g$ and $z$ bands of Pan-STARRS to those of the catalogues, which have been used by \citet{2022A&A...661A...3S},
allows us to apply the same criteria as \citet{2022A&A...661A...3S} for our sources (see Sect.\,\ref{algorithm}).

\subsection{Catalogues of AGNs and galaxies}
\label{AGN-cata}
The Million Quasars (Milliquas) catalogue \citep{2021yCat.7290....0F} is the main catalogue, which was used to flag the background sources. Almost all known AGN have been already included in these catalogues, but to include probably neglected candidates we run source matching on the  catalogues of following studies as well: the SWIFT AGN and Cluster Survey \citep{2015ApJS..218....8D}, the identification of 1.4 Million AGNs in the MIR using WISE Data \citep{2015ApJS..221...12S}, and the identifications of AGNs from the WISE, 2MASS, and \rosat\, All-Sky Surveys \citep{2012ApJ...751...52E}.

\section{Algorithm for the source classification}
\label{algorithm-sec}
We searched for the optical and IR counterparts of the X-ray sources (within their 3$\sigma$ positional error) using the NWAY code \citet{2018MNRAS.473.4937S} (See Sect.\,\ref{nway}).  However,  when the number density  of counterparts in a  catalogue is high in a field (i.e. as in the Galactic disk), a single X-ray source may have multiple counterparts. Therefore, the match probability reduces and consequently, it is very hard to find a significant counterpart for an X-ray source. The situation is even more dramatic in cases of faint X-ray sources with a large positional error and this is similar to what we have seen for many of the \erosita\ sources. The weighted average of the positional error of  point-like sources \citep[flagged with EXT=0][]{2024A&A...682A..34M}  of our catalogue is 4.5". As Fig.\,\ref{num-counter} shows, 37$\%$ of sources have at least two WISE counterparts and  67$\%$ have 3 \gaia\,counterparts. Moreover, the majority of these counterparts cannot be considered as a significant match according to the match probability of NWAY. Results of the cross-match of X-ray sources in the field of CMa~OD shows that only $\sim$25$\%$ of sources have a significant WISE counterpart ({\selectfont p-i>90$\%$}) and only $\sim$0.7$\%$ of sources have a  significant \gaia\, counterpart. To find the likely counterpart and, thus, a classification of the unknown and faint sources, we developed an algorithm for the classification of X-ray sources.
 
We focussed on the study of fields including a nearby galaxy, where the contribution of X-ray background sources and Galactic foreground sources has to be removed before we can identify the candidates of X-rays sources belonging to the galaxy. This algorithm is based on the available multi-wavelength data and criteria, which have been already applied for the data, plus the X-ray properties of the \erosita\, sources. As mentioned in Sects.\,\ref{infra-sec} and \,\ref{optical-sec}, we used the two most updated all-sky catalogues in the optical\,(\gaia) and IR\,(WISE). Using these surveys offers two benefits. First, they cover the whole sky and this allows our algorithm to be applied to other classification studies as well. Second, these surveys have already been frequently used for the population study of different types of sources based on their available bands or colours; thus, the results of these previous studies can be applied to our study for the classification of the \erosita\, sources. Our algorithm does not define a limit with respect to accepting a counterpart based on its match probability; instead, it starts with the most probable optical and/or IR counterparts (as we call it the first counterpart); afterwards, if the first counterpart does not satisfy the classification criteria, it checks the next probable counterpart (if it exists within the 3$\sigma$ positional circle error of the source) based on considerations explained below. 

\begin{figure*}[!ht]
\includegraphics[trim={0.cm 0.2cm 0.0cm 0.7cm}, width=0.45\textwidth]{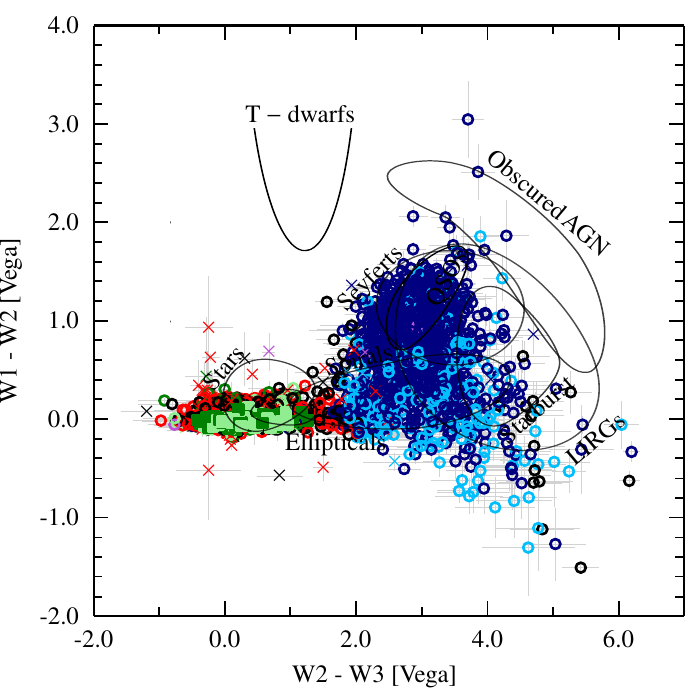}
\includegraphics[trim={0.cm 0.2cm 0.0cm 0.7cm},clip, width=0.55\textwidth]{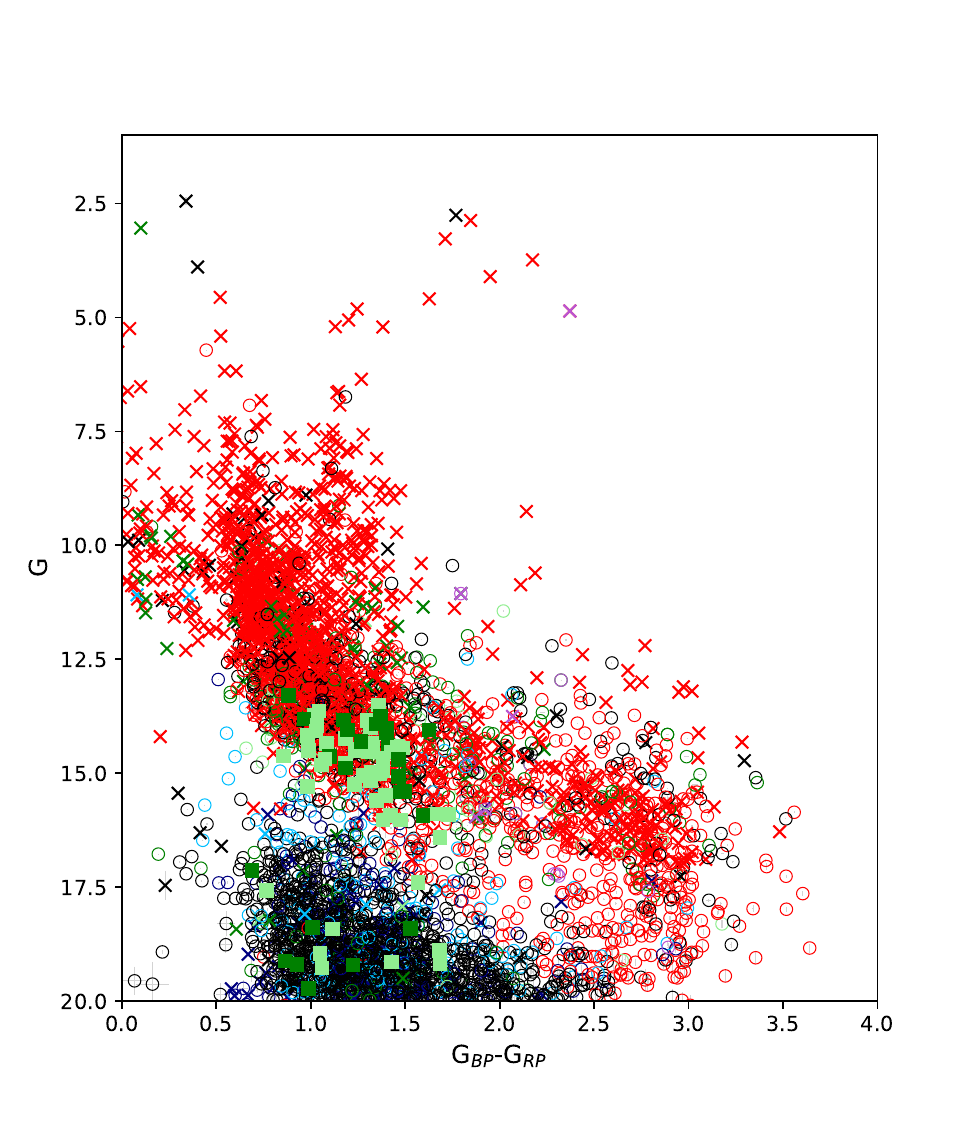}
\caption{{\bf } Colour-colour diagram of IR WISE counterparts of X-ray sources in the field of CMa~OD (left). The region of different types of sources are shown based on the study of \citet{2010AJ....140.1868W}. {\bf } Colour-magnitude diagram of \gaia\,counterparts of the X-ray sources in the field of CMa~OD (right). The background sources, Galactic foreground sources, stellar objects, and candidate members of CMa~OD, plus their candidates are shown in different colours and symbols. Colour red indicates Galactic foreground stars and stellar systems, orange show foreground object candidates. Classified background sources are shown in blue and the candidates in light blue. Sources in green are stellar objects and light green are stellar object candidates. The symbiotic candidates are shown in purple. The sources without classification are shown in black. Moreover, the known sources with entries in SIMBAD are shown with crosses and sources, which are classified through this work are marked with circles. Stellar objects that are considered members of CMa~OD are shown with filled squares.
\label{wise-plot}}
\end{figure*}

Figure\,\ref{algorithm} shows the schematic of the algorithm. An unknown X-ray source, which completes the pipeline, will be classified as a background source, a Galactic foreground star or binary, or a stellar object; sources that are a star candidate according to the WISE colours, but the \gaia\,information is not significant enough to clarify whether they belong to the MW or they are extra Galactic sources (e.g. they belong to the CMa~OD) or it stays unclassified. Figures\,\ref{wise-plot} and\,\ref{x-opt-plot} show the results of the algorithm pipeline, as explained in Sect.\,\ref{algorithm-sec}. In the very first step, the pipeline checks whether an X-ray source is a known object according to SIMBAD or the  most updated available  catalogues of AGNs or galaxies. The used catalogues of background sources are listed in Sect.\,\ref{AGN-cata}. Overall, 1901 sources have been identified as known sources in SIMBAD and 507 sources exist in AGN or quasar catalogues. If a source is unknown, in the second step, the algorithm checks the first optical and first IR counterparts of an X-ray source are identical\,(i.e. if they are located at the same position within the positional uncertainty of \gaia\,and WISE). For many X-ray sources, the most probable optical and IR counterparts are not identical (see Fig.\,\ref{num-counter}). 

\subsection{Counterparts of a source}
After recognising the first optical and IR counterparts, there are four possibilities:
\label{counter-sect}

\begin{figure}[!htp]
\includegraphics[clip, trim={0.0cm  0.0cm  0.0cm  0.cm},width=0.50\textwidth]{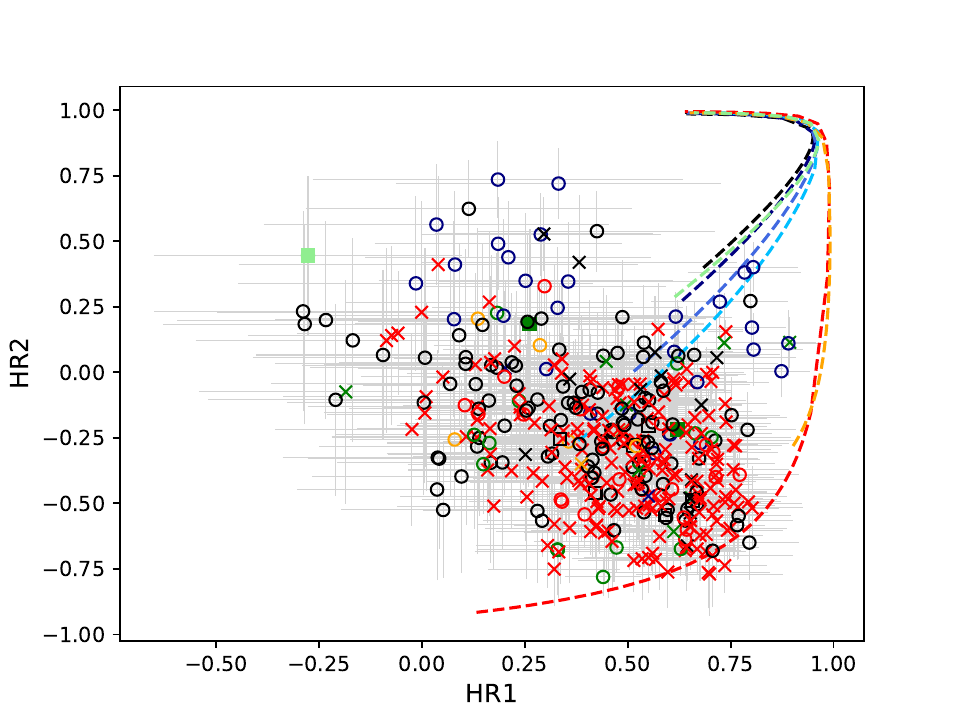}
\includegraphics[clip, trim={0.0cm  0.0cm  0.0cm  0.cm},width=0.50\textwidth]{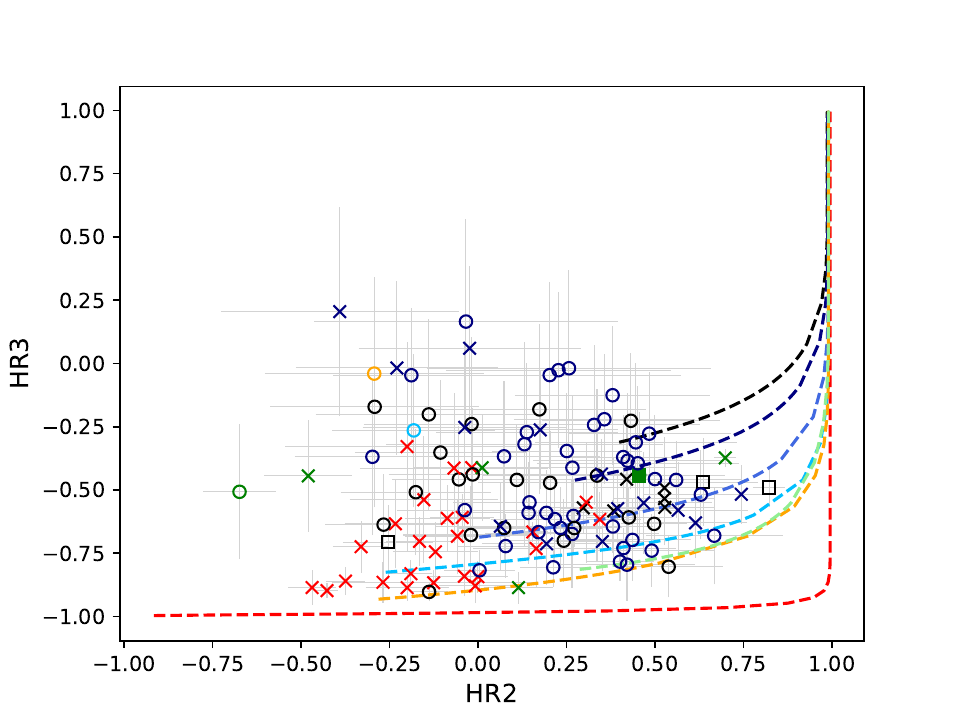}
\includegraphics[clip, trim={0.0cm  0.0cm  0.0cm  0.cm},width=0.50\textwidth]{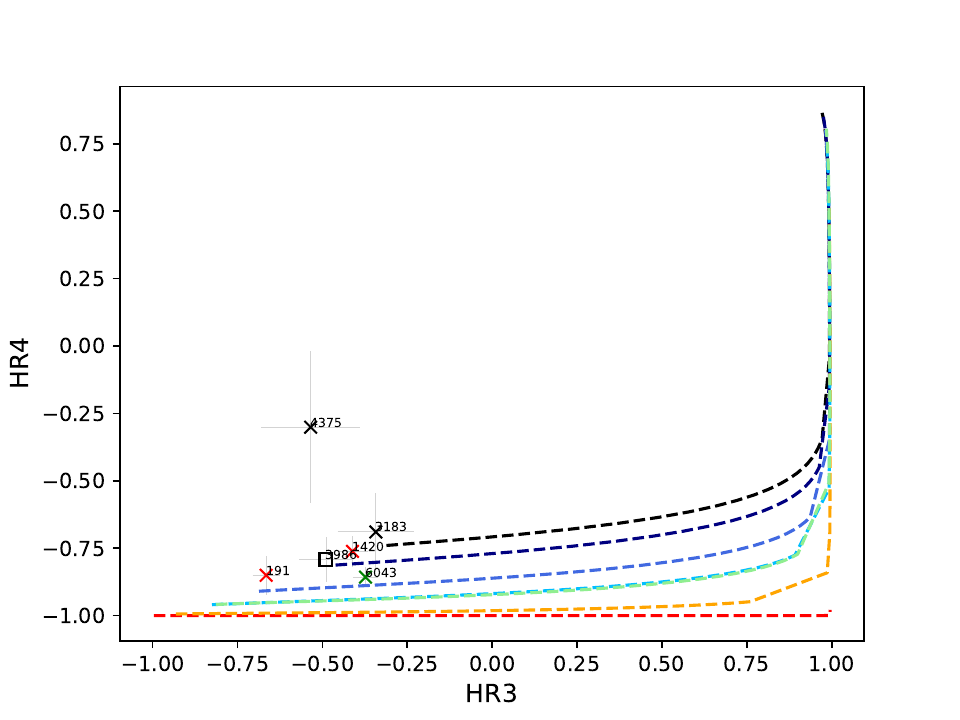}
\hspace{1cm}
\caption{HR diagrams. The lines display the hardness ratios calculated for different spectral models. Black, navy, and blue lines show a power-law model with a photon index of $\Gamma$=1, 2, and 3,  respectively. Red, yellow, and green lines show the apec model \citep{2001ApJ...556L..91S} for a thermal plasma with a temperature of $kT=$ 0.2, 1.0, and 2.0~keV, respectively.  The upper plot shows sources in the foreground, which therefore have very low absorbing foreground column density $N_{\rm H}$; they appear to be softer than the others. The symbols and colours are the same as in Fig.\,\ref{wise-plot}. Additionally, those sources, which have no counterpart are shown with black squares.  
\label{hrs-plot}}
\end{figure}

{\it \fontfamily{phv}\selectfont No counterpart:} the X-ray source has neither an optical nor an IR counterpart within its $3\sigma$ positional error. These sources are considered for further investigations if, for instance, any ultraviolet (UV) or radio counterpart exists. Otherwise the study is limited to the spectral analysis of the X-ray source. Spectral analysis or the hardness ratios (HRs, see Sect.\,\ref{hr-sec}) are helpful to find candidates with soft X-ray emission (e.g. supernova remnants) or hard sources (e.g. XRBs with an obscured companion). Here, we have 95 sources without any counterparts. None of these sources had a matching radio or UV source, while a significant HR (See Sect.\,\ref{hr-sec}) was available for 10 of these sources. They are labeled with black squares in the HR plots\,(see Figs.\,\ref{hrs-plot} and\,\ref{sss-plot}) and are further discussed in Sect.\,\ref{diss-sec}.

{\it \fontfamily{phv}\selectfont Only \gaia\,counterpart:} The most probable counterpart is only a single \gaia\,source. 
First, this \gaia\,counterpart (as the most probable counterpart) is checked for the parallax information. However, if the \gaia\,counterpart is confirmed as a foreground Galactic source, still we cannot be sure that the X-ray emission is from a Galactic source. Since the most numerous and bright X-ray sources are AGNs, we can expect that  there still might be an AGN  within the X-ray positional error (as second or third counterpart), which is partially or fully responsible for the X-ray emission. For this reason, a foreground Galactic object is only accepted after checking the second and third optical or IR counterparts. An X-ray source with both a foreground and a background counterpart would remains unclassified or marked as a foreground Galactic source 
(see Fig.\,\ref{algorithm}).

{\it \fontfamily{phv}\selectfont  Only WISE counterpart:} The most probable counterpart is
only a single WISE source, without any \gaia\,counterpart.
If the WISE counterpart is significant in terms of its bands, it is checked if it is a candidate for  a background source or a stellar object  (see Sect.\,\ref{crit-sect}). 

{\it \fontfamily{phv}\selectfont Matched WISE and \gaia~counterparts:} The first \gaia\,counterpart is within the positional uncertainty of the first WISE counterpart. First, the source is checked to see whether it is a background object (see Sect.\,\ref{crit-sect}). If it is not a background source, it is checked to see whether it fulfills the condition for Galactic sources.  If a source is neither a background nor a Galactic source, the last step is a check to see whether it fulfills the conditions for stellar objects. A stellar object can either be a Galactic source or belong to the nearby galaxy (here CMa~OD). To show that a source belongs to CMa~OD, additional IR or optical criteria are needed,  as explained in Sect.\,\ref{diss-sec}.

If the first counterpart(s) of a source remains without any classification through the algorithm, the algorithm starts from the beginning to check whether the second optical or IR counterparts can be classified and, in the next step, the third counterparts.

\subsection{Criteria for the classification}
\label{crit-sect}
The criteria for the classification of different types of X-ray sources are explained in the following.

 { \it \fontfamily{phv}\selectfont Background objects}: There are deep studies in both IR and optical aimed at classifying AGNs combined with X-ray studies. We applied the following three criteria to classify a source as a background source. If a source satisfies one (or more) of these conditions it is considered as a background object. 
1) \citet{2010AJ....140.1868W} have shown that background objects are usually expected to be red (i.e. $W2-W3$>1.5). This condition is checked only for the sources, which have been significantly detected in $W2$ and $W3$. Figure\,\ref{wise-plot} shows the separation of background sources and stellar objects based on the WISE colours; 2) On the other hand, if we have a significant optical counterpart for an X-ray source in the Pan-STARRS1 Survey, the deep study of the \erosita\, Final Equatorial-Depth Survey (eFEDS) by \citet{2022A&A...661A...3S} shows that extra-galactic sources are characterised by $z-W1-0.8*(g-r)+1.2>0$ and $W1+1.625*log(F_{(0.5-2.)})+6.101>0$. This set of criteria are useful for sources that do not have a significant $W2$ or $W3$ detection; therefore, the first criterion cannot be applied. However, these criteria  cannot be applied as the main criteria in our work, because the number of sources that have a Pan-STARRS1 counterpart is limited to $\sim$15$\%$ of the sources, which is much lower than the counterparts in the WISE catalogue (see left plot of Fig.\,\ref{num-counter}). Overall, $\sim90\%$ of sources have at least one WISE counterpart.

{\it \fontfamily{phv}\selectfont Galactic single stars and binary systems:} To identify the Galactic objects, the parallax measurement of \gaia\,is crucial.  We have taken into account the significance of the \gaia\,parallax measurement following these three criteria suggested by the \gaia\,collaboration for DR3: {\fontfamily{qcr}\selectfont ruwe}\,(re-normalised unit weight error)\,<\,1.4, {\fontfamily{qcr}\selectfont ipd$\_$frac$\_$multi$\_$peak}\,(percentage of CCD transits where additional images were seen)\,$\leq$\,2,  and {\fontfamily{qcr}\selectfont ipd$\_$gof$\_$harmonic$\_$amplitude\,}\,(scan angle dependent variation of goodness of fit for the image fitting)\,<\,0.1 \citep[see also][]{2021A&A...649A...5F}. However, some more recent works \citep[e.g.][]{2022A&A...667A..74A} show that the {\fontfamily{qcr}\selectfont ruwe} parameter is not fully reliable for filtering unreliable parallax measurements. For this reason, in cases where significant WISE measurements are available, we have also taken into account the condition of stellar objects for Galactic sources as well (i.e. $W2-W3$<1.5). If the IR colour and the \gaia\,parallax measurement of a counterpart are not compatible, the X-ray source remains unclassified. Moreover, the aforementioned conditions of \gaia\,parameters do not cover Galactic binary systems. To identify Galactic binary systems, we cross-matched our source list with the catalogue of \citet{2021MNRAS.506.2269E}, which lists binary systems up to a distance of $\sim$1\,kpc.

{\it \fontfamily{phv}\selectfont Stellar objects:} When Galactic sources are excluded, we expect that the IR counterpart of an X-ray source, which is an accreting binary system in a nearby galaxy (e.g. CMa~OD,  satisfies the condition of stellar objects. According to \citet{2010AJ....140.1868W}, these object have a colour of $W2-W3$<1.5. 

The details of the magnitudes of WISE and \gaia\,counterparts, which have been considered in source classification using the algorithm are presented in Table~\ref{counterparts-cata}.
\begin{figure*}[!htp]
\includegraphics[trim={0.cm 0.0cm 0.0cm 0.7cm},clip, width=0.5\textwidth]{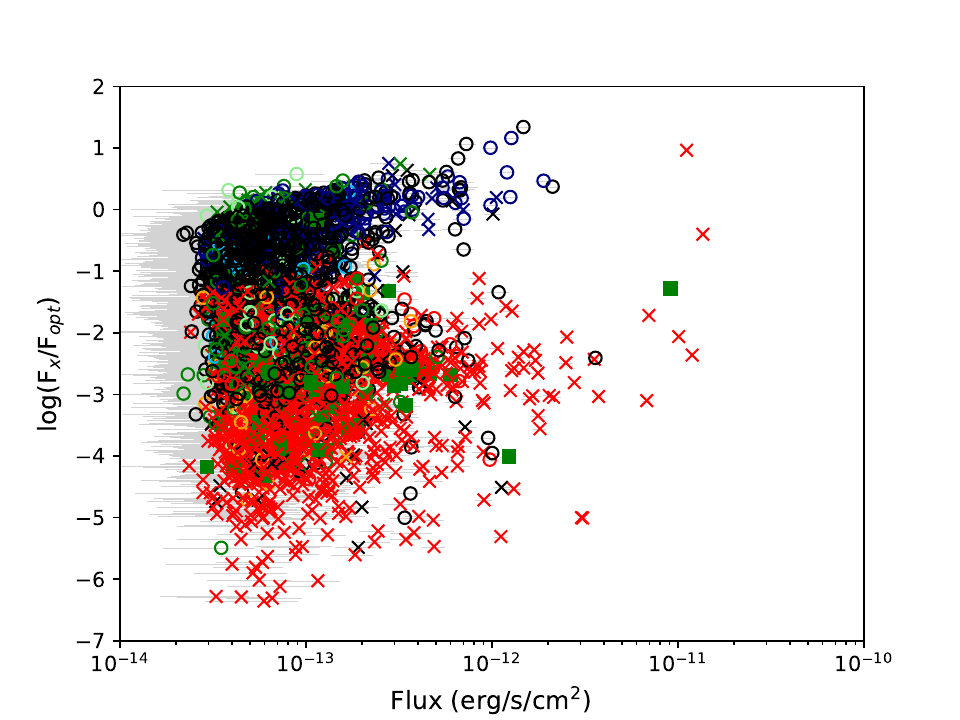}
\includegraphics[trim={0.cm 0.2cm 0.0cm 0.7cm},clip, width=0.5\textwidth]{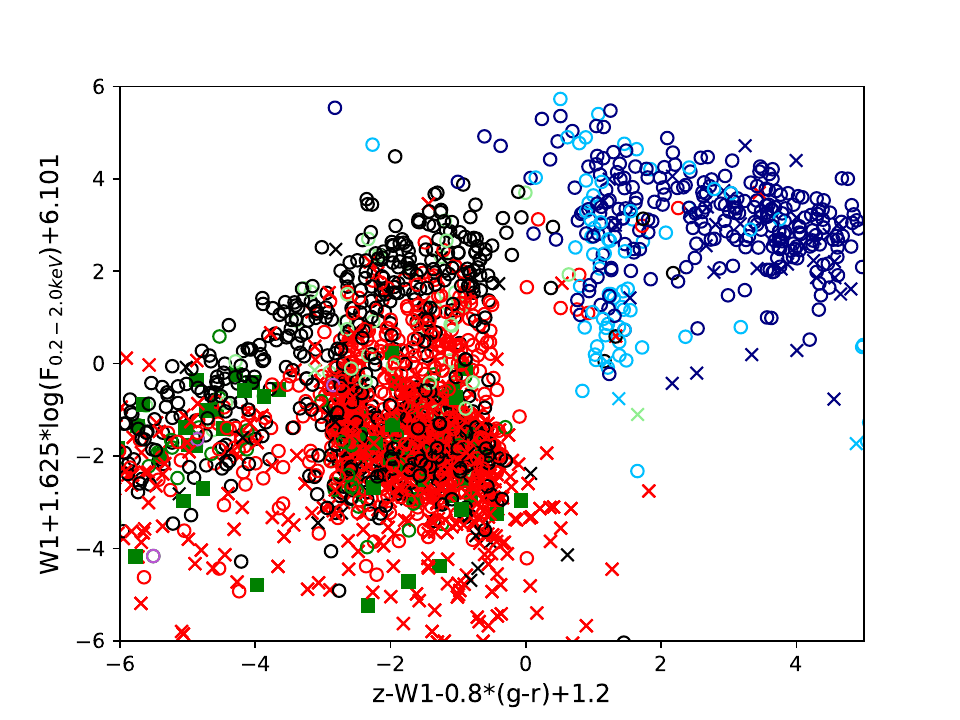}
\caption{{\bf } Logarithmic X-ray 
  to optical flux ratio versus the 0.2--2.3\,keV X-ray flux (left). {\bf } Right panel shows the criteria based on optical magnitudes\,($g$, $r$, $z$), IR magnitude ($W1$), and X-ray flux to distinguish background object from  Galactic sources,  as suggested by \citet{2022A&A...661A...3S}. The symbols and colours are the same as in Fig.\,\ref{wise-plot}.\label{x-opt-plot}}
\end{figure*}

\section{X-ray analysis}
\label{xray-sect}
For the classification of the X-ray sources, we also used the information of the ratio between the X-ray flux and the optical magnitude and the spectral properties given as hardness ratios (HRs).

\subsection {X-ray to optical flux ratio}
\label{x-opt-sect}
The X-ray-to-optical-flux ratio, calculated as the ratio between the X-ray flux and the optical magnitude of the counterpart, is a useful tool for identifying, for instance, optically bright stars versus\ X-ray bright accreting compact objects.  We plotted X-ray to optical flux ratio over the X-ray flux in Fig.\,\ref{x-opt-plot} (left). The following equation shows how ${\rm log}\bigg(\frac{F_ \text{X}}{F_\text{opt}}\bigg)$ was calculated \citep[see][]{1988ApJ...326..680M}, adapted to the \gaia\,bands:
\begin{equation}
{\rm log}\bigg(\frac{F_ \text{X}}{F_\text{opt}}\bigg)={{\rm log}_{10}(F_\text{X})}+\frac{G_{BP}+G_{RP}}{2\times2.5}+5.37,
\end{equation}
where $F_\text{X}$ is the X-ray flux, and $G_{BP}$ and $G_{RP}$ are the \gaia\,magnitudes of the optical counterpart associated with the X-ray source.

In addition, \citet{2022A&A...661A...3S} provided  criteria to distinguish between the background sources\,(AGNs) and Galactic sources using the X-ray flux and the IR and optical magnitudes (see Sect.\,\ref{algorithm-sec}). The right plot of Fig.\,\ref{x-opt-plot} shows that sources, which were classified in our work based on  PanSTARRS1 and WISE,  satisfy the conditions of \citet{2022A&A...661A...3S} as well. 

\subsection {Hardness ratios}
\label{hr-sec}
The count rate of majority of  X-ray sources in the field of our study is not enough for the spectral analysis.
There are many X-ray sources (including stellar objects or unclassified sources) that are too faint for a spectral analysis. Therefore, we calculated the hardness ratios, which give information about the spectral properties of the sources. Hardness ratios were calculated from the rate (counts/s) of sources in the five available energy bands of the eRASS1 catalogue, namely, 0.2--0.5~keV, 0.5,--1.0~keV, 1.0--2.0~keV, 2.0--5.0~keV, and 5.0--8.0~keV. To increase the accuracy, we considered an HR measurement significant only when the detection likelihood in the two corresponding energy bands is higher than 6\,(i.e.~>3$\sigma$). The HRs and HR errors are calculated as: 
\begin{equation} 
HR_\mathrm{i}=\frac{B_\mathrm{i+1}-B_\mathrm{i}}{B_\mathrm{i+1}+B_\mathrm{i}} ~~\mathrm{and}~~ EHR_i=2\frac{\sqrt{(B_\mathrm{i+1}EB_\mathrm{i})^2+(B_\mathrm{i}EB_\mathrm{i+1})^2}} {(B_\mathrm{i+1}+B_\mathrm{i})^2}, 
\end{equation}
respectively, where $B_\mathrm{i}$ is the count rate and $EB_\mathrm{i}$ is corresponding error in the energy band,~$i$. 

The hardness ratio plots HR2 versus HR1 and HR3 versus HR2 (Fig.\,\ref{hrs-plot}) show that the population of foreground Galactic stars are mainly consistent with models of collisionally
ionized thermal plasma \citep[apec,][]{2001ApJ...556L..91S} in comparison to background sources, which are mainly consistent with power-low models. There are a few very hard sources in the HR4 versus HR3 diagram (Fig.\,\ref{hrs-plot}). The details of these objects are discussed in Sect.\,\ref{diss-sec}.

\section{Discussion}
\label{diss-sec}

The aim of this work is to provide a primary view about the population of X-ray sources, which belong to the CMa~OD. We  did not present any X-ray spectral analysis of bright background AGNs or Galactic systems, as these sources are addressed by other teams in the \erosita\ consortium.
Here, bright X-ray sources in the field of CMa~OD are considered for further studies to verify whether they can be classified as systems belonging to CMa~OD (based on our algorithm); otherwise, they remain unclassified.
\begin{figure}[!ht]
\includegraphics[trim={0.cm 0.2cm 0.0cm 0.7cm},clip, width=0.5\textwidth]{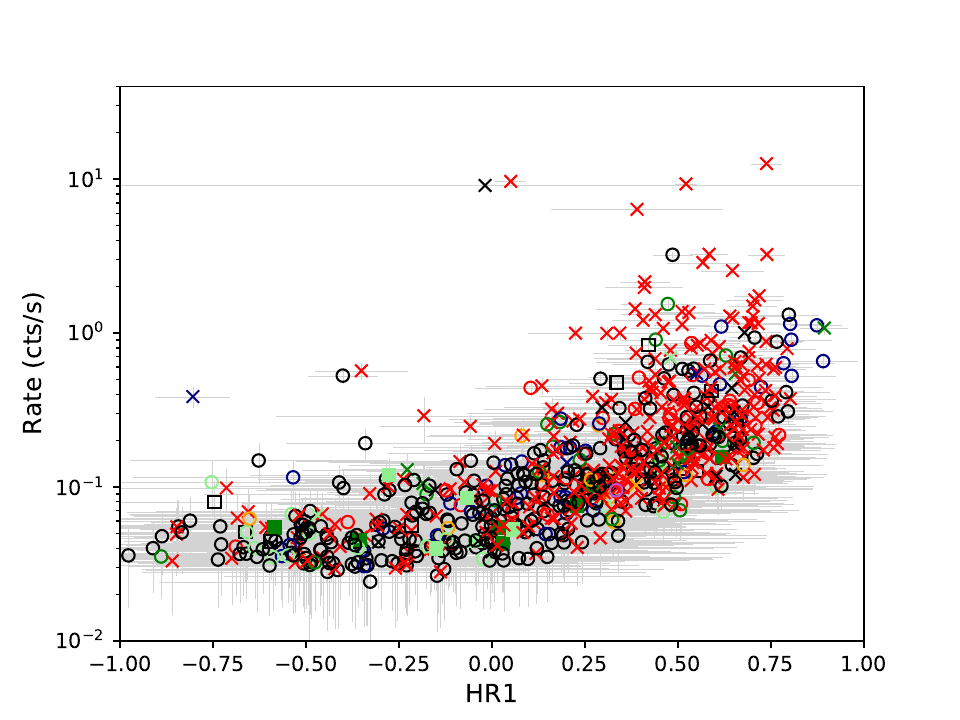}
\caption{Count rate of the X-ray sources versus  HR1. Colours and symbols are the same as in Fig.\,\ref{wise-plot}. Additionally, those sources, which have no counterpart are shown with black squares. \label{sss-plot}}
\end{figure}

\subsection{Members of CMa~OD}
\label{cma-member-sec}
\begin{figure}[!ht]
\includegraphics[trim={0.cm 0.2cm 0.0cm 0.7cm},clip, width=0.5\textwidth]{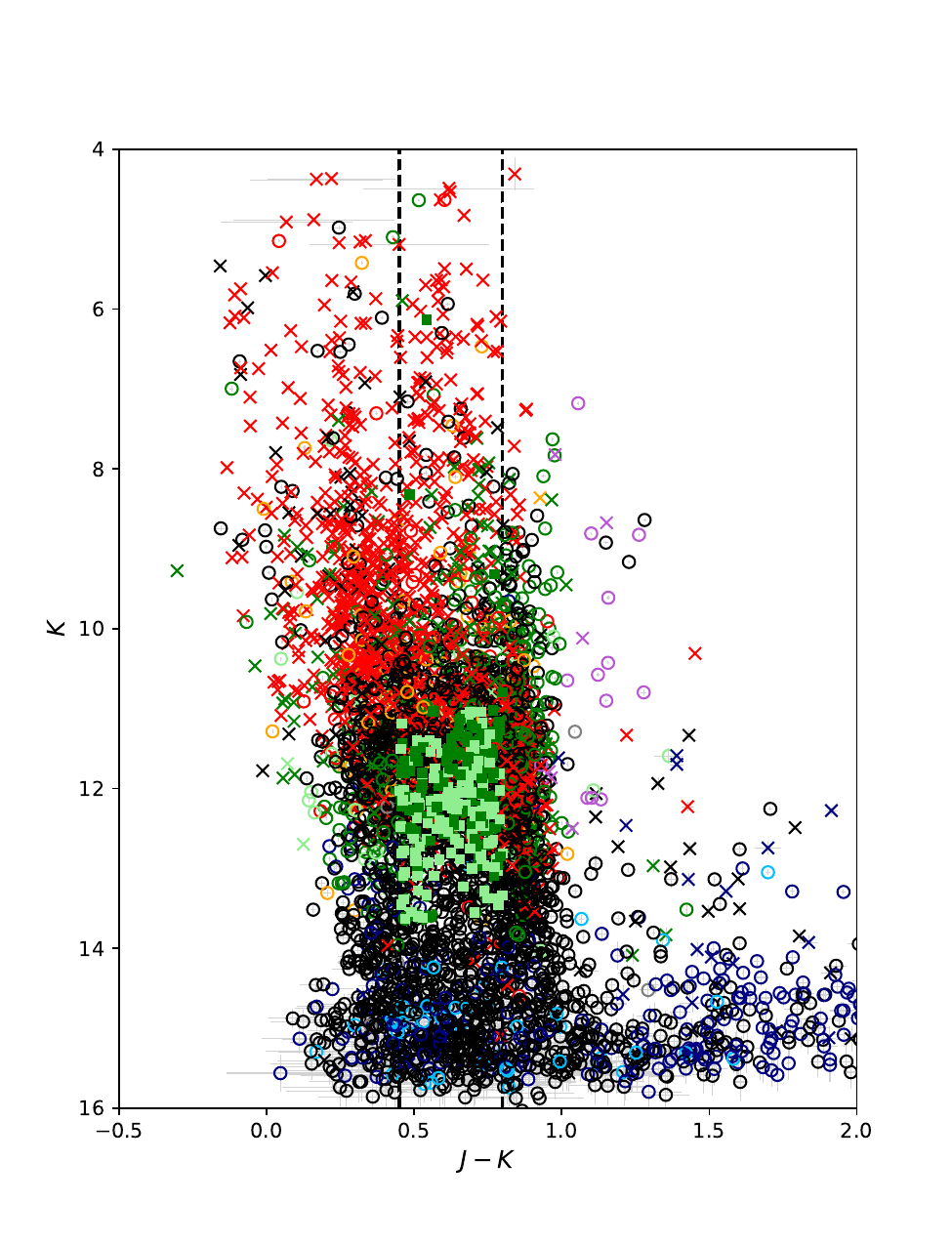}
\caption{ Colour-magnitude
  diagram of 2MASS IR counterparts. A noticeable population of
  X-ray sources has IR counterparts, which have $K$ and $J-K$ typical of M-giants in CMa~OD\citep[$0.45<J-K<0.8$, dashed lines,
  and $11\lesssim K \lesssim 14$;][]{2006MNRAS.366..865B}. The colours and symbols are the same as in Fig.\,\ref{wise-plot}.}
  \label{2mass-plot}
\end{figure}

In our classification process, we found 95 sources without any optical or IR counterpart and 435 sources, which are stellar systems. Due to the higher distance and thus the lack of accurate parallax measurements for stars in CMa~OD  it is difficult to distinguish the members of CMa~OD from faint Galactic stars. However, the IR red-clump population of CMa~OD shows a colour of 0.45 $\leq J-K \leq$ 0.75 and the main population has a magnitude of 12.3 $<K<$ 13.7 \citep{2006MNRAS.366..865B}. We  checked this colour and magnitude using the data of 2MASS (see Fig.\,\ref{2mass-plot}) for sources classified as stellar objects. We classified these sources as CMa~OD members if they fulfil these criteria along with ML > 10, whereas they were  only determined to be as member candidates when ML < 10. We classified 100 sources as members of CMa~OD and
235 as  member candidates of CMa~OD. They are shown with green filled squares in all plots. 

\begin{figure}[!htp]
\begin{center}
    \includegraphics[angle=270, trim={1.5cm 0.0cm 0.0cm 2.cm},clip,width=0.50\textwidth]{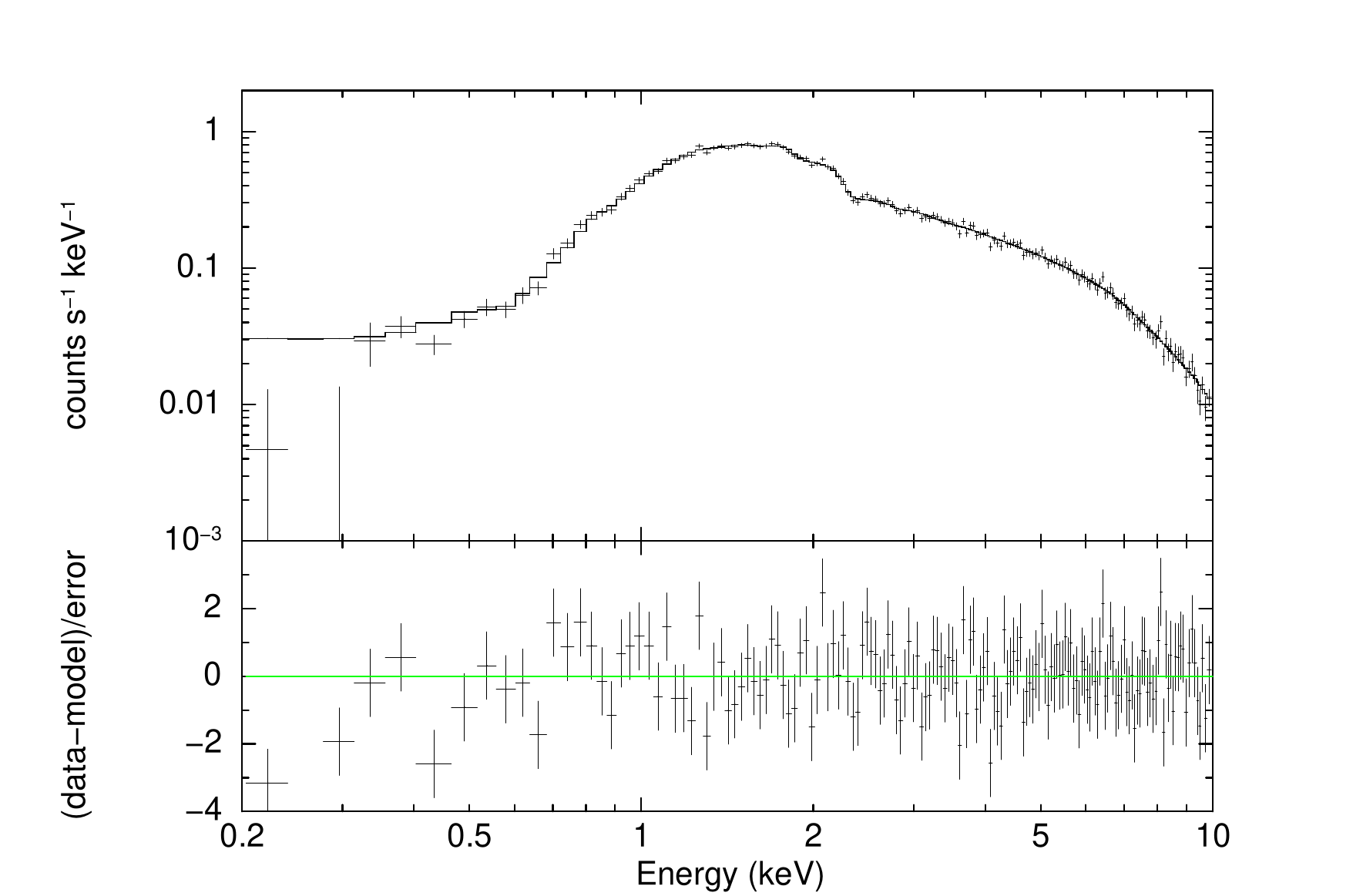}
    \caption{X-ray spectrum of \xmm\, EPIC-pn of SRGEt\,J071522.1-191609.}
    \label{xmm-spec}
\end{center}
\end{figure}

Moreover we checked the X-ray HRs and X-ray flux of the brighter sources. The HR plots (Fig.\,\ref{hrs-plot}) show that there are a few stellar systems (green circles) in HR1-HR2 and HR2-HR3 diagrams, which mainly show soft X-ray emission <\,2.0~keV. The HR plot of higher energies (HR3-HR4) shows a few bright and hard X-ray sources, which are two known bright quasars (source No.\,4375 and No.\,3183) and two known Galactic cataclysmic variables (CVs, source No.\,191 and source No.\,1420, see Sect.\,\ref{final-cata}). Moreover, source No.\,6043 is a well-known Be\,HMXB V*\,V441\,Pup (also known as 3A\,0726--260 or 4U\,0728--25), which has a Be~star companion with a distance of $\sim$ 8~kpc \citep{2023A&A...674A...1G}. Thus, the source seems to belong to CMa~OD and there is a possibility that it is part of the population of blue giant stars in CMa~OD \citep{2021MNRAS.501.1690C}. The last source in HR3-HR4 plot is source No.\,3986\,(SRGEt\,J071522.1-191609), which has been discovered by \erosita\, via the Near Real Time Analysis (NRTA) when the position entered the \erosita\, field of view on 2020-04-19\,(MJD 58958.06) \citep[][]{2020ATel13657....1G}. The \erosita\ flux is $(1.3\pm0.4)\times10^{-11}$ erg\,s$^{-1}$\,cm$^{2}$ assuming an absorbed power-law model with photon index ($\Gamma= 2.1\pm0.6$) for the source. The source was not detected by \rosat\, and considering its upper limit flux in \rosat\ data ($(4.2\pm0.4)\times10^{-13}$ erg\,s$^{-1}$\,cm$^{2}$), it has brightened by a factor of 12 over the last $\sim$30 years.
\citet{2020ATel13716....1V}  analysed  follow-up observations of this source carried out with \swift\, and the Very Large Array\,(VLA). The spectrum of \swift/XRT is fitted with an absorbed power-law model with a photon index of $\Gamma=\,1.9\pm0.25$, resulting in an unabsorbed flux of $(2.\pm0.3)10^{-11}$~erg\,s$^{-1}$\,cm$^{-2}$. The VLA observation confirms a radio counterpart for the source. According to the spectral index of $\alpha$=0.18$\pm$0.25, calculated from  fluxes of 193$\pm$22 $\mu$Jy and 176$\pm$7 $\mu$Jy at 4.5\,(GHz) and 7.5\,(GHz) bands, respectively (where the flux density is $S_{\nu}\propto \nu^{\alpha}$), \citet{2020ATel13716....1V} suggested that the spectrum is similar to that of a compact jet of X-ray binaries. Moreover, they  concluded that assuming a distance of $<20.$\,kpc for the source, its X-ray luminosity matches either a black hole X-ray binary or a radio-bright neutron star X-ray binary. Furthermore, by assuming a distance of 7 kpc, the X-ray and radio luminosities support an X-ray binary nature for a low-luminosity transient \citep{2020ATel13716....1V}. In addition, available optical catalogues show no counterpart for the source, while the optical follow-up observation of iTelescope in Siding Spring, Australia confirms an optical counterpart for the source with a magnitude of $R=19.1\pm0.1$ \citep[][]{2020ATel13669....1K}. We note that the source was observed by \xmm\, in 2021. Figure\,\ref{xmm-spec} shows the spectrum of \xmm\, EPIC-pn camera. The spectrum is fitted well with an absorbed (black-body+power\,law) model. The \nh\, value of $7.72(\pm0.30)\times10^{21}$\,cm$^{-2}$ is comparable to the Galactic column density in the direction of SRGEt\,J071522.1-191609, therefore, no intrinsic absorption for the system has been confirmed in the spectrum. The black-body has a temperature of $1.59^{+0.42}_{-0.71}$ keV and the power\,law has a photon index of $\Gamma=1.81\pm0.10$. The spectrum indicates an unabsorbed flux of $(1.68\pm0.2)\times10^{-11}$~erg\,s$^{-1}$\,cm$^{-2}$ \,(consistent with the \swift\, flux) and a luminosity of $1.27\times10^{35}$ ~erg\,s$^{-1}$ assuming a distance of 7\,kpc.  The light curve of SRGEt\,J071522.1-191609 shows no dimming, flare, or significant variability over the \xmm\,observing time. A Z$^{2}$-analysis \citep{1983A&A...128..245B} using the first and the second harmonic values of $n=1$, 2 on the barycentrically corrected event files to search for periodic signals in the period range 0.114\,s--$\sim$25\,ks  to a period corresponding to the length of observation ($\sim$25 ks) and a Lomb-Scargle periodogram fail to reveal a significant periodicity for the source. 
Assuming a typical luminosity for transient LMXBs \citep[i.e, $10^{32}$--$10^{34}$; e.g.][]{2022A&A...661A..35S}, the source can be a member of CMa~OD with a distance of $\geq$ 7.~kpc.


So far, neither super-soft sources nor supernova remnants (SNRs) have been identified in CMa~OD. Figure\,\ref{sss-plot} presents the count rate of the X-ray sources over the HR1.  The bright and soft X-ray sources are typically Galactic sources (in red). There are a few very soft X-ray sources (HR1<0) that might belong to CMa~OD (green). However, they are too faint for a spectral analysis using \erosita\, data and follow-up X-ray observations are required.

\begin{figure*}[!ht]
\includegraphics[trim={0.cm 0.0cm 0.0cm 0.0cm},clip, width=0.45\textwidth]{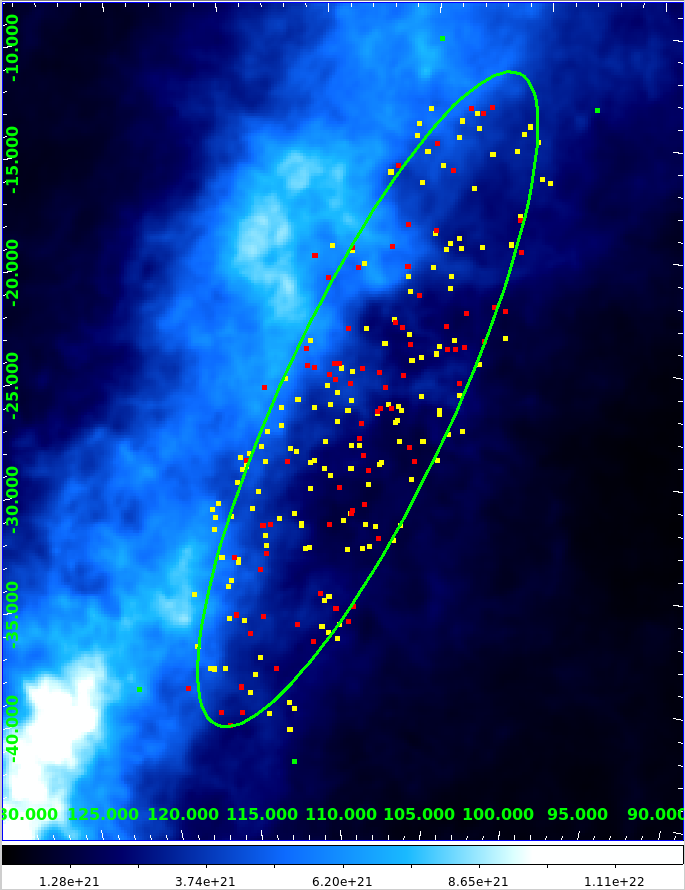}
\includegraphics[trim={0.cm 0.0cm 1.0cm 0.0cm},clip, width=0.55\textwidth]{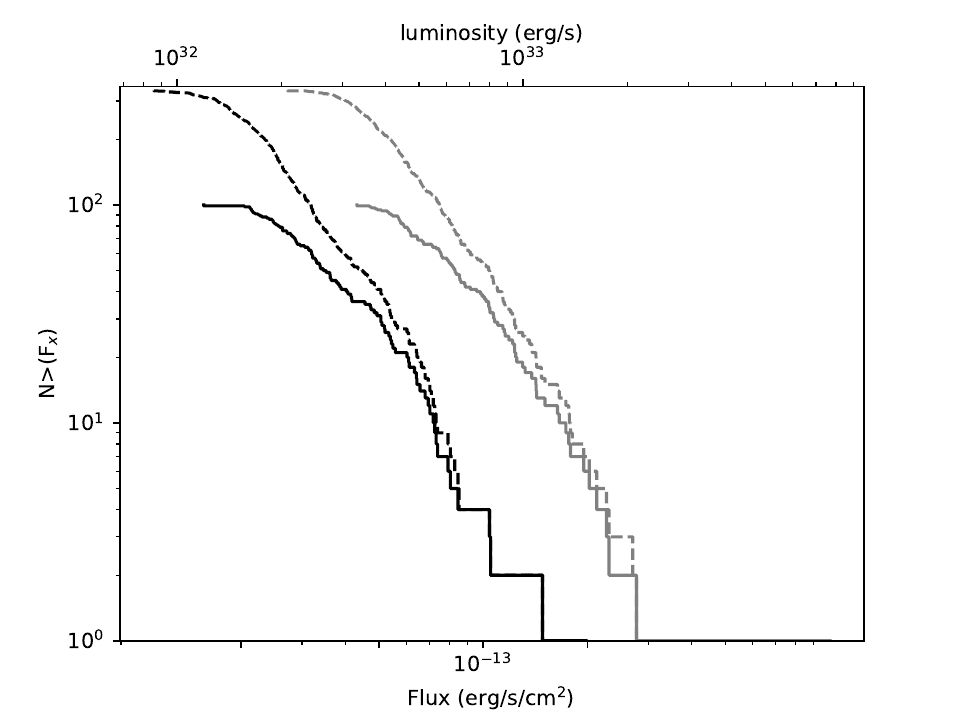}
\caption{ {\bf }HI column density ({$N_{H}$}) map in the CMa~OD field from the HI4PI survey (left). The red dots show the position of X-ray sources classified as CMa~OD members and the yellow dots are the CMa~OD member candidates.  The XLF of sources, which are classified as CMa~OD members or candidates, are shown on the right. All sources\,(members and candidate members of CMa~OD) are plotted by dashed lines and only the members by solid lines. The  black and gray lines show the absorbed and unabsorbed fluxes, respectively. \label{xlf-plot}}
\end{figure*}

Moreover, we have also identified symbiotic star candidates using the criteria based on the machine learning method of \citet{2019MNRAS.483.5077A}. With respect to the sources for which WISE and 2MASS counterparts were found and which have been classified as stellar objects, we have checked the following IR colours. \citet{2019MNRAS.483.5077A} showed that the majority of symbiotic stars appear to have $J-H>0.78$ and only a small fraction of S-type symbiotics behave differently. The second criterion is  $K-W3<1.18$, which separates the symbiotic stars significantly from the other types of sources. The second criterion might not apply for  dusty symbiotic stars. In this case, there are two other criteria for the colours: $H-W2>3.80$ and $W1-W4<4.72$. We applied all the above conditions to classify symbiotic stars in the field of CMa~OD. We found 34 sources as symbiotic star candidates, which are marked with a purple colour in all plots. None of these sources were bright enough to perform an X-ray spectral analysis, so they belong to either CMa~OD or to the Milky Way.  However, due to the lack of significant parallax measurements, their distances remain unknown. 

\subsection{XLF}
In a population of low-mass stars as in CMa~OD, we would expect  the majority of X-ray sources to form  the three classes of AWDs, transient LMXBs, and active binaries. To understand the luminosity of the X-ray sources in CMa~OD, we have to consider that it is the closest dSph to the Milky Way and is located behind the Galactic disk; therefore, not only  is the apparent size of this galaxy large, but the foreground column density is not homogeneous over the entire dSph (see the left plot of Fig.\,\ref{xlf-plot}).
We  used the {$N_{H}$} measurements of the \citet{2016A&A...594A.116H}. However, for the absorption of X-ray sources we  need an estimate of the molecular hydrogen column density\,({$N_{H2}$}) as well to calculate the total column density of $N_{H}$+$2N_{H2}$. We estimated the  {$N_{H2}$} using the equation provided by \citep{2013MNRAS.431..394W}, which allows us to estimate the {$N_{H2}$} using {$N_{H}$} and the extinction values (E($B-V$)).
We calculated the flux of each source in the energy range of 0.2--2.0 keV assuming a power-law model, with a photon index of $\Gamma=2;$  this is the value typically considered for the contribution of unknown sources to the XLFs that are most likely to be AGNs \citep[e.g.][]{2009A&A...497..635C}.
For each source, the absorbed and the unabsorbed flux were estimated using the {\selectfont cflux} model of {\selectfont Xspec\,(V.12.13.0)}. Figure\,\ref{xlf-plot} shows the result of the XLF of CMa~OD members and member candidates. It presents a population of X-ray sources with an unabsorbed luminosity between $2\times10^{32}$--$10^{34}$~erg\,s$^{-1}$, which is expected to be mainly a population of AWDs or transient LMXBs. Only $\sim$10 sources have a luminosity $>10^{33}$~erg\,s$^{-1}$, which are candidates for either symbiotic stars or transient X-ray sources. 
We do not have a population of active binaries as  eRASS1 is not sensitive enough to detect sources with luminosities of $<10^{32}$~erg\,s$^{-1}$ at the distance presented by CMa~OD.

\section{Summary}
 With the goal  of classifying  X-ray sources detected with \erosita, we performed a multi-wavelength analysis of the counterparts of X-ray sources in the field of CMa~OD. The majority of the faint sources exhibit a large X-ray positional error, which leads to the association of multiple optical or IR counterparts for each source. Therefore, we have developed an algorithm to select the most probable class for a source in a systematic way based on the available information of optical and IR counterparts. 

From a total number of 8311 sources, 1901 have entries in SIMBAD with classification (see Table.\,\ref{catalogue-x-ray}). 507 sources have been already identified as background sources in other catalogues. In this study, we have successfully identified 946 new background sources and 274 candidates. We classified 1029 sources as Galactic foreground objects, 20 sources as Galactic foreground candidates, 435 sources as stellar objects (which may either be Galactic sources or  members of CMa~OD), and 189 sources as stellar object candidates. 
Out of the stellar objects, 100 sources are considered to be members of CMa~OD and 235 to be candidate members of CMa~OD. They are expected to be mainly AWD systems according to their X-ray luminosities. Finally, 2936 sources remained unclassified. 

Due to the short exposure time of eRASS1 (average of 250\,s in the field of CMa~OD) that results in poor photon statistics, the study of the spectral properties of the sources was limited to the HRs.  There was no SNR, nor any super-soft source found in the CMa~OD. Moreover, a few hard sources were detected, which are all known sources. As discussed in Sect.\,\ref{cma-member-sec}, one source is a known HMXB and another is a new transient LMXB candidate. Our results show a significant number of low-luminosity X-ray sources that are candidates for different types of AWDs. None of these sources were bright enough for their spectra to be studied in detail. The X-ray sources in CMa~OD are ideal targets for follow-up X-ray studies and surveys given the  likelihood that they are transient X-ray sources.
\begin{acknowledgements}
This work was supported by the Deutsche Forschungsgemeinschaft through the project
SA 2131/15-1.
This work is based on data from \erosita, the soft X-ray instrument aboard
SRG, a joint Russian-German science mission supported by the Russian Space
Agency (Roskosmos), in the interests of the Russian Academy of Sciences
represented by its Space Research Institute (IKI), and the Deutsches Zentrum für
Luft- und Raumfahrt (DLR). The SRG spacecraft was built by Lavochkin
Association (NPOL) and its subcontractors, and is operated by NPOL with
support from the Max Planck Institute for Extraterrestrial Physics (MPE).
The development and construction of the \erosita\, X-ray instrument was led by
MPE, with contributions from the Dr. Karl Remeis Observatory Bamberg \& ECAP
(FAU Erlangen-Nuernberg), the University of Hamburg Observatory, the Leibniz
Institute for Astrophysics Potsdam (AIP), and the Institute for Astronomy and
Astrophysics of the University of Tübingen, with the support of DLR and the Max
Planck Society. The Argelander Institute for Astronomy of the University of Bonn
and the Ludwig Maximilians Universität Munich also participated in the science
preparation for \erosita.
This  research  has  made  use  of  the  SIMBAD  and  VIZIER  database,  operated at  CDS,  Strasbourg,  France,  and  of  the  NASA/IPAC  Extra-galactic  Database\,(NED),  which  is  operated  by  the  Jet  Propulsion  Laboratory,  California  Institute  of  Technology,  under  contract  with  the  National  Aeronautics  and  Space Administration. This publication makes use of data products from the Wide field Infrared Survey Explorer, which is a joint project of the University of California, Los Angeles, and the Jet Propulsion Laboratory/California Institute of Technology, funded by the National Aeronautics and Space Administration. This publication has made use of data products from the Two Micron All Sky Survey, which is a joint project of the University of Massachusetts and the Infrared Processing  and  Analysis  Center,  funded  by  the  National  Aeronautics  and  Space Administration  and  the  National  Science  Foundation.  This research has made use of SAO Image DS9, developed by Smithsonian Astrophysical Observatory.
 The Digitized Sky Surveys were produced at the Space Telescope Science Institute under U.S. Government grant NAG W-2166. The images of these surveys are based on photographic data obtained using the Oschin Schmidt Telescope on Palomar Mountain and the UK Schmidt Telescope. The plates were processed into the present compressed digital form with the permission of these institutions.
The National Geographic Society - Palomar Observatory Sky Atlas (POSS-I) was made by the California Institute of Technology with grants from the National Geographic Society.
The Second Palomar Observatory Sky Survey (POSS-II) was made by the California Institute of Technology with funds from the National Science Foundation, the National Geographic Society, the Sloan Foundation, the Samuel Oschin Foundation, and the Eastman Kodak Corporation.
The Oschin Schmidt Telescope is operated by the California Institute of Technology and Palomar Observatory.
The UK Schmidt Telescope was operated by the Royal Observatory Edinburgh, with funding from the UK Science and Engineering Research Council (later the UK Particle Physics and Astronomy Research Council), until 1988 June, and thereafter by the Anglo-Australian Observatory. The blue plates of the southern Sky Atlas and its Equatorial Extension (together known as the SERC-J), as well as the Equatorial Red (ER), and the Second Epoch [red] Survey (SES) were all taken with the UK Schmidt.

\end{acknowledgements}

\bibliographystyle{aa} 
\bibliography{bibtex}

\begin{thebibliography}{54}
\expandafter\ifx\csname natexlab\endcsname\relax\def\natexlab#1{#1}\fi

\bibitem[{{Aihara} {et~al.}(2018){Aihara}, {Armstrong}, {Bickerton}, {Bosch},
  {Coupon}, {Furusawa}, {Hayashi}, {Ikeda}, {Kamata}, {Karoji}, {Kawanomoto},
  {Koike}, {Komiyama}, {Lang}, {Lupton}, {Mineo}, {Miyatake}, {Miyazaki},
  {Morokuma}, {Obuchi}, {Oishi}, {Okura}, {Price}, {Takata}, {Tanaka},
  {Tanaka}, {Tanaka}, {Uchida}, {Uraguchi}, {Utsumi}, {Wang}, {Yamada},
  {Yamanoi}, {Yasuda}, {Arimoto}, {Chiba}, {Finet}, {Fujimori}, {Fujimoto},
  {Furusawa}, {Goto}, {Goulding}, {Gunn}, {Harikane}, {Hattori}, {Hayashi},
  {He{\l}miniak}, {Higuchi}, {Hikage}, {Ho}, {Hsieh}, {Huang}, {Huang},
  {Imanishi}, {Iwata}, {Jaelani}, {Jian}, {Kashikawa}, {Katayama}, {Kojima},
  {Konno}, {Koshida}, {Kusakabe}, {Leauthaud}, {Lee}, {Lin}, {Lin},
  {Mandelbaum}, {Matsuoka}, {Medezinski}, {Miyama}, {Momose}, {More}, {More},
  {Mukae}, {Murata}, {Murayama}, {Nagao}, {Nakata}, {Niida}, {Niikura},
  {Nishizawa}, {Oguri}, {Okabe}, {Ono}, {Onodera}, {Onoue}, {Ouchi}, {Pyo},
  {Shibuya}, {Shimasaku}, {Simet}, {Speagle}, {Spergel}, {Strauss}, {Sugahara},
  {Sugiyama}, {Suto}, {Suzuki}, {Tait}, {Takada}, {Terai}, {Toba}, {Turner},
  {Uchiyama}, {Umetsu}, {Urata}, {Usuda}, {Yeh}, \&
  {Yuma}}]{2018PASJ...70S...8A}
{Aihara}, H., {Armstrong}, R., {Bickerton}, S., {et~al.} 2018, \pasj, 70, S8

\bibitem[{{Akras} {et~al.}(2019){Akras}, {Leal-Ferreira}, {Guzman-Ramirez}, \&
  {Ramos-Larios}}]{2019MNRAS.483.5077A}
{Akras}, S., {Leal-Ferreira}, M.~L., {Guzman-Ramirez}, L., \& {Ramos-Larios},
  G. 2019, \mnras, 483, 5077

\bibitem[{{Andriantsaralaza} {et~al.}(2022){Andriantsaralaza}, {Ramstedt},
  {Vlemmings}, \& {De Beck}}]{2022A&A...667A..74A}
{Andriantsaralaza}, M., {Ramstedt}, S., {Vlemmings}, W.~H.~T., \& {De Beck}, E.
  2022, \aap, 667, A74

\bibitem[{{Bailer-Jones} {et~al.}(2021){Bailer-Jones}, {Rybizki}, {Fouesneau},
  {Demleitner}, \& {Andrae}}]{2021AJ....161..147B}
{Bailer-Jones}, C.~A.~L., {Rybizki}, J., {Fouesneau}, M., {Demleitner}, M., \&
  {Andrae}, R. 2021, \aj, 161, 147

\bibitem[{{Bellazzini} {et~al.}(2002){Bellazzini}, {Ferraro}, {Origlia},
  {Pancino}, {Monaco}, \& {Oliva}}]{2002AJ....124.3222B}
{Bellazzini}, M., {Ferraro}, F.~R., {Origlia}, L., {et~al.} 2002, \aj, 124,
  3222

\bibitem[{{Bellazzini} {et~al.}(2006){Bellazzini}, {Ibata}, {Martin}, {Lewis},
  {Conn}, \& {Irwin}}]{2006MNRAS.366..865B}
{Bellazzini}, M., {Ibata}, R., {Martin}, N., {et~al.} 2006, \mnras, 366, 865

\bibitem[{{Brunner} {et~al.}(2022){Brunner}, {Liu}, {Lamer}, {Georgakakis},
  {Merloni}, {Brusa}, {Bulbul}, {Dennerl}, {Friedrich}, {Liu}, {Maitra},
  {Nandra}, {Ramos-Ceja}, {Sanders}, {Stewart}, {Boller}, {Buchner}, {Clerc},
  {Comparat}, {Dwelly}, {Eckert}, {Finoguenov}, {Freyberg}, {Ghirardini},
  {Gueguen}, {Haberl}, {Kreykenbohm}, {Krumpe}, {Osterhage}, {Pacaud},
  {Predehl}, {Reiprich}, {Robrade}, {Salvato}, {Santangelo}, {Schrabback},
  {Schwope}, \& {Wilms}}]{2022A&A...661A...1B}
{Brunner}, H., {Liu}, T., {Lamer}, G., {et~al.} 2022, \aap, 661, A1

\bibitem[{{Buccheri} {et~al.}(1983){Buccheri}, {Bennett}, {Bignami}, {Bloemen},
  {Boriakoff}, {Caraveo}, {Hermsen}, {Kanbach}, {Manchester}, {Masnou},
  {Mayer-Hasselwander}, {{\"O}zel}, {Paul}, {Sacco}, {Scarsi}, \&
  {Strong}}]{1983A&A...128..245B}
{Buccheri}, R., {Bennett}, K., {Bignami}, G.~F., {et~al.} 1983, \aap, 128, 245

\bibitem[{{Cappelluti} {et~al.}(2009){Cappelluti}, {Brusa}, {Hasinger},
  {Comastri}, {Zamorani}, {Finoguenov}, {Gilli}, {Puccetti}, {Miyaji},
  {Salvato}, {Vignali}, {Aldcroft}, {B{\"o}hringer}, {Brunner}, {Civano},
  {Elvis}, {Fiore}, {Fruscione}, {Griffiths}, {Guzzo}, {Iovino}, {Koekemoer},
  {Mainieri}, {Scoville}, {Shopbell}, {Silverman}, \&
  {Urry}}]{2009A&A...497..635C}
{Cappelluti}, N., {Brusa}, M., {Hasinger}, G., {et~al.} 2009, \aap, 497, 635

\bibitem[{{Carballo-Bello} {et~al.}(2021){Carballo-Bello},
  {Mart{\'\i}nez-Delgado}, {Corral-Santana}, {Alfaro}, {Navarrete}, {Vivas}, \&
  {Catelan}}]{2021MNRAS.501.1690C}
{Carballo-Bello}, J.~A., {Mart{\'\i}nez-Delgado}, D., {Corral-Santana}, J.~M.,
  {et~al.} 2021, \mnras, 501, 1690

\bibitem[{{Cutri} \& {et al.}(2014)}]{2014yCat.2328....0C}
{Cutri}, R.~M. \& {et al.} 2014, VizieR Online Data Catalog, II/328

\bibitem[{{Cutri} {et~al.}(2003){Cutri}, {Skrutskie}, {van Dyk}, {Beichman},
  {Carpenter}, {Chester}, {Cambresy}, {Evans}, {Fowler}, {Gizis}, {Howard},
  {Huchra}, {Jarrett}, {Kopan}, {Kirkpatrick}, {Light}, {Marsh}, {McCallon},
  {Schneider}, {Stiening}, {Sykes}, {Weinberg}, {Wheaton}, {Wheelock}, \&
  {Zacarias}}]{2003yCat.2246....0C}
{Cutri}, R.~M., {Skrutskie}, M.~F., {van Dyk}, S., {et~al.} 2003, VizieR Online
  Data Catalog, II/246

\bibitem[{{Dai} {et~al.}(2015){Dai}, {Griffin}, {Kochanek}, {Nugent}, \&
  {Bregman}}]{2015ApJS..218....8D}
{Dai}, X., {Griffin}, R.~D., {Kochanek}, C.~S., {Nugent}, J.~M., \& {Bregman},
  J.~N. 2015, \apjs, 218, 8

\bibitem[{{de Boer} {et~al.}(2018){de Boer}, {Belokurov}, \&
  {Koposov}}]{2018MNRAS.473..647D}
{de Boer}, T.~J.~L., {Belokurov}, V., \& {Koposov}, S.~E. 2018, \mnras, 473,
  647

\bibitem[{{de Jong} {et~al.}(2007){de Jong}, {Butler}, {Rix}, {Dolphin}, \&
  {Mart{\'\i}nez-Delgado}}]{2007ApJ...662..259D}
{de Jong}, J.~T.~A., {Butler}, D.~J., {Rix}, H.~W., {Dolphin}, A.~E., \&
  {Mart{\'\i}nez-Delgado}, D. 2007, \apj, 662, 259

\bibitem[{{Dey} {et~al.}(2019){Dey}, {Schlegel}, {Lang}, {Blum}, {Burleigh},
  {Fan}, {Findlay}, {Finkbeiner}, {Herrera}, {Juneau}, {Landriau}, {Levi},
  {McGreer}, {Meisner}, {Myers}, {Moustakas}, {Nugent}, {Patej}, {Schlafly},
  {Walker}, {Valdes}, {Weaver}, {Y{\`e}che}, {Zou}, {Zhou}, {Abareshi},
  {Abbott}, {Abolfathi}, {Aguilera}, {Alam}, {Allen}, {Alvarez}, {Annis},
  {Ansarinejad}, {Aubert}, {Beechert}, {Bell}, {BenZvi}, {Beutler}, {Bielby},
  {Bolton}, {Brice{\~n}o}, {Buckley-Geer}, {Butler}, {Calamida}, {Carlberg},
  {Carter}, {Casas}, {Castander}, {Choi}, {Comparat}, {Cukanovaite}, {Delubac},
  {DeVries}, {Dey}, {Dhungana}, {Dickinson}, {Ding}, {Donaldson}, {Duan},
  {Duckworth}, {Eftekharzadeh}, {Eisenstein}, {Etourneau}, {Fagrelius},
  {Farihi}, {Fitzpatrick}, {Font-Ribera}, {Fulmer}, {G{\"a}nsicke},
  {Gaztanaga}, {George}, {Gerdes}, {Gontcho}, {Gorgoni}, {Green}, {Guy},
  {Harmer}, {Hernandez}, {Honscheid}, {Huang}, {James}, {Jannuzi}, {Jiang},
  {Joyce}, {Karcher}, {Karkar}, {Kehoe}, {Kneib}, {Kueter-Young}, {Lan},
  {Lauer}, {Le Guillou}, {Le Van Suu}, {Lee}, {Lesser}, {Perreault Levasseur},
  {Li}, {Mann}, {Marshall}, {Mart{\'\i}nez-V{\'a}zquez}, {Martini}, {du Mas des
  Bourboux}, {McManus}, {Meier}, {M{\'e}nard}, {Metcalfe},
  {Mu{\~n}oz-Guti{\'e}rrez}, {Najita}, {Napier}, {Narayan}, {Newman}, {Nie},
  {Nord}, {Norman}, {Olsen}, {Paat}, {Palanque-Delabrouille}, {Peng},
  {Poppett}, {Poremba}, {Prakash}, {Rabinowitz}, {Raichoor}, {Rezaie},
  {Robertson}, {Roe}, {Ross}, {Ross}, {Rudnick}, {Safonova}, {Saha},
  {S{\'a}nchez}, {Savary}, {Schweiker}, {Scott}, {Seo}, {Shan}, {Silva},
  {Slepian}, {Soto}, {Sprayberry}, {Staten}, {Stillman}, {Stupak}, {Summers},
  {Sien Tie}, {Tirado}, {Vargas-Maga{\~n}a}, {Vivas}, {Wechsler}, {Williams},
  {Yang}, {Yang}, {Yapici}, {Zaritsky}, {Zenteno}, {Zhang}, {Zhang}, {Zhou}, \&
  {Zhou}}]{2019AJ....157..168D}
{Dey}, A., {Schlegel}, D.~J., {Lang}, D., {et~al.} 2019, \aj, 157, 168

\bibitem[{{Ducati}(2002)}]{2002yCat.2237....0D}
{Ducati}, J.~R. 2002, VizieR Online Data Catalog

\bibitem[{{Edelson} \& {Malkan}(2012)}]{2012ApJ...751...52E}
{Edelson}, R. \& {Malkan}, M. 2012, \apj, 751, 52

\bibitem[{{El-Badry} {et~al.}(2021){El-Badry}, {Rix}, \&
  {Heintz}}]{2021MNRAS.506.2269E}
{El-Badry}, K., {Rix}, H.-W., \& {Heintz}, T.~M. 2021, \mnras, 506, 2269

\bibitem[{{Fabricius} {et~al.}(2021){Fabricius}, {Luri}, {Arenou}, {Babusiaux},
  {Helmi}, {Muraveva}, {Reyl{\'e}}, {Spoto}, {Vallenari}, {Antoja}, {Balbinot},
  {Barache}, {Bauchet}, {Bragaglia}, {Busonero}, {Cantat-Gaudin}, {Carrasco},
  {Diakit{\'e}}, {Fabrizio}, {Figueras}, {Garcia-Gutierrez}, {Garofalo},
  {Jordi}, {Kervella}, {Khanna}, {Leclerc}, {Licata}, {Lambert}, {Marrese},
  {Masip}, {Ramos}, {Robichon}, {Robin}, {Romero-G{\'o}mez}, {Rubele}, \&
  {Weiler}}]{2021A&A...649A...5F}
{Fabricius}, C., {Luri}, X., {Arenou}, F., {et~al.} 2021, \aap, 649, A5

\bibitem[{{Flesch}(2021)}]{2021yCat.7290....0F}
{Flesch}, E.~W. 2021, VizieR Online Data Catalog, VII/290

\bibitem[{{Forbes} \& {Bridges}(2010)}]{2010MNRAS.404.1203F}
{Forbes}, D.~A. \& {Bridges}, T. 2010, \mnras, 404, 1203

\bibitem[{{Frebel} {et~al.}(2010){Frebel}, {Kirby}, \&
  {Simon}}]{2010Natur.464...72F}
{Frebel}, A., {Kirby}, E.~N., \& {Simon}, J.~D. 2010, \nat, 464, 72

\bibitem[{{Gaia Collaboration} {et~al.}(2023){Gaia Collaboration}, {Vallenari},
  {Brown}, {Prusti}, {de Bruijne}, {Arenou}, {Babusiaux}, {Biermann},
  {Creevey}, {Ducourant}, {Evans}, {Eyer}, {Guerra}, {Hutton}, {Jordi},
  {Klioner}, {Lammers}, {Lindegren}, {Luri}, {Mignard}, {Panem}, {Pourbaix},
  {Randich}, {Sartoretti}, {Soubiran}, {Tanga}, {Walton}, {Bailer-Jones},
  {Bastian}, {Drimmel}, {Jansen}, {Katz}, {Lattanzi}, {van Leeuwen}, {Bakker},
  {Cacciari}, {Casta{\~n}eda}, {De Angeli}, {Fabricius}, {Fouesneau},
  {Fr{\'e}mat}, {Galluccio}, {Guerrier}, {Heiter}, {Masana}, {Messineo},
  {Mowlavi}, {Nicolas}, {Nienartowicz}, {Pailler}, {Panuzzo}, {Riclet}, {Roux},
  {Seabroke}, {Sordo}, {Th{\'e}venin}, {Gracia-Abril}, {Portell}, {Teyssier},
  {Altmann}, {Andrae}, {Audard}, {Bellas-Velidis}, {Benson}, {Berthier},
  {Blomme}, {Burgess}, {Busonero}, {Busso}, {C{\'a}novas}, {Carry}, {Cellino},
  {Cheek}, {Clementini}, {Damerdji}, {Davidson}, {de Teodoro}, {Nu{\~n}ez
  Campos}, {Delchambre}, {Dell'Oro}, {Esquej}, {Fern{\'a}ndez-Hern{\'a}ndez},
  {Fraile}, {Garabato}, {Garc{\'\i}a-Lario}, {Gosset}, {Haigron}, {Halbwachs},
  {Hambly}, {Harrison}, {Hern{\'a}ndez}, {Hestroffer}, {Hodgkin}, {Holl},
  {Jan{\ss}en}, {Jevardat de Fombelle}, {Jordan}, {Krone-Martins}, {Lanzafame},
  {L{\"o}ffler}, {Marchal}, {Marrese}, {Moitinho}, {Muinonen}, {Osborne},
  {Pancino}, {Pauwels}, {Recio-Blanco}, {Reyl{\'e}}, {Riello}, {Rimoldini},
  {Roegiers}, {Rybizki}, {Sarro}, {Siopis}, {Smith}, {Sozzetti}, {Utrilla},
  {van Leeuwen}, {Abbas}, {{\'A}brah{\'a}m}, {Abreu Aramburu}, {Aerts},
  {Aguado}, {Ajaj}, {Aldea-Montero}, {Altavilla}, {{\'A}lvarez}, {Alves},
  {Anders}, {Anderson}, {Anglada Varela}, {Antoja}, {Baines}, {Baker},
  {Balaguer-N{\'u}{\~n}ez}, {Balbinot}, {Balog}, {Barache}, {Barbato},
  {Barros}, {Barstow}, {Bartolom{\'e}}, {Bassilana}, {Bauchet}, {Becciani},
  {Bellazzini}, {Berihuete}, {Bernet}, {Bertone}, {Bianchi}, {Binnenfeld},
  {Blanco-Cuaresma}, {Blazere}, {Boch}, {Bombrun}, {Bossini}, {Bouquillon},
  {Bragaglia}, {Bramante}, {Breedt}, {Bressan}, {Brouillet}, {Brugaletta},
  {Bucciarelli}, {Burlacu}, {Butkevich}, {Buzzi}, {Caffau}, {Cancelliere},
  {Cantat-Gaudin}, {Carballo}, {Carlucci}, {Carnerero}, {Carrasco},
  {Casamiquela}, {Castellani}, {Castro-Ginard}, {Chaoul}, {Charlot}, {Chemin},
  {Chiaramida}, {Chiavassa}, {Chornay}, {Comoretto}, {Contursi}, {Cooper},
  {Cornez}, {Cowell}, {Crifo}, {Cropper}, {Crosta}, {Crowley}, {Dafonte},
  {Dapergolas}, {David}, {David}, {de Laverny}, {De Luise}, {De March}, {De
  Ridder}, {de Souza}, {de Torres}, {del Peloso}, {del Pozo}, {Delbo},
  {Delgado}, {Delisle}, {Demouchy}, {Dharmawardena}, {Di Matteo}, {Diakite},
  {Diener}, {Distefano}, {Dolding}, {Edvardsson}, {Enke}, {Fabre}, {Fabrizio},
  {Faigler}, {Fedorets}, {Fernique}, {Fienga}, {Figueras}, {Fournier},
  {Fouron}, {Fragkoudi}, {Gai}, {Garcia-Gutierrez}, {Garcia-Reinaldos},
  {Garc{\'\i}a-Torres}, {Garofalo}, {Gavel}, {Gavras}, {Gerlach}, {Geyer},
  {Giacobbe}, {Gilmore}, {Girona}, {Giuffrida}, {Gomel}, {Gomez},
  {Gonz{\'a}lez-N{\'u}{\~n}ez}, {Gonz{\'a}lez-Santamar{\'\i}a},
  {Gonz{\'a}lez-Vidal}, {Granvik}, {Guillout}, {Guiraud},
  {Guti{\'e}rrez-S{\'a}nchez}, {Guy}, {Hatzidimitriou}, {Hauser}, {Haywood},
  {Helmer}, {Helmi}, {Sarmiento}, {Hidalgo}, {Hilger}, {H{\l}adczuk}, {Hobbs},
  {Holland}, {Huckle}, {Jardine}, {Jasniewicz}, {Jean-Antoine Piccolo},
  {Jim{\'e}nez-Arranz}, {Jorissen}, {Juaristi Campillo}, {Julbe}, {Karbevska},
  {Kervella}, {Khanna}, {Kontizas}, {Kordopatis}, {Korn}, {K{\'o}sp{\'a}l},
  {Kostrzewa-Rutkowska}, {Kruszy{\'n}ska}, {Kun}, {Laizeau}, {Lambert},
  {Lanza}, {Lasne}, {Le Campion}, {Lebreton}, {Lebzelter}, {Leccia}, {Leclerc},
  {Lecoeur-Taibi}, {Liao}, {Licata}, {Lindstr{\o}m}, {Lister}, {Livanou},
  {Lobel}, {Lorca}, {Loup}, {Madrero Pardo}, {Magdaleno Romeo}, {Managau},
  {Mann}, {Manteiga}, {Marchant}, {Marconi}, {Marcos}, {Marcos Santos},
  {Mar{\'\i}n Pina}, {Marinoni}, {Marocco}, {Marshall}, {Martin Polo},
  {Mart{\'\i}n-Fleitas}, {Marton}, {Mary}, {Masip}, {Massari},
  {Mastrobuono-Battisti}, {Mazeh}, {McMillan}, {Messina}, {Michalik}, {Millar},
  {Mints}, {Molina}, {Molinaro}, {Moln{\'a}r}, {Monari}, {Mongui{\'o}},
  {Montegriffo}, {Montero}, {Mor}, {Mora}, {Morbidelli}, {Morel}, {Morris},
  {Muraveva}, {Murphy}, {Musella}, {Nagy}, {Noval}, {Oca{\~n}a}, {Ogden},
  {Ordenovic}, {Osinde}, {Pagani}, {Pagano}, {Palaversa}, {Palicio},
  {Pallas-Quintela}, {Panahi}, {Payne-Wardenaar}, {Pe{\~n}alosa Esteller},
  {Penttil{\"a}}, {Pichon}, {Piersimoni}, {Pineau}, {Plachy}, {Plum}, {Poggio},
  {Pr{\v{s}}a}, {Pulone}, {Racero}, {Ragaini}, {Rainer}, {Raiteri}, {Rambaux},
  {Ramos}, {Ramos-Lerate}, {Re Fiorentin}, {Regibo}, {Richards}, {Rios Diaz},
  {Ripepi}, {Riva}, {Rix}, {Rixon}, {Robichon}, {Robin}, {Robin}, {Roelens},
  {Rogues}, {Rohrbasser}, {Romero-G{\'o}mez}, {Rowell}, {Royer}, {Ruz Mieres},
  {Rybicki}, {Sadowski}, {S{\'a}ez N{\'u}{\~n}ez}, {Sagrist{\`a} Sell{\'e}s},
  {Sahlmann}, {Salguero}, {Samaras}, {Sanchez Gimenez}, {Sanna},
  {Santove{\~n}a}, {Sarasso}, {Schultheis}, {Sciacca}, {Segol}, {Segovia},
  {S{\'e}gransan}, {Semeux}, {Shahaf}, {Siddiqui}, {Siebert}, {Siltala},
  {Silvelo}, {Slezak}, {Slezak}, {Smart}, {Snaith}, {Solano}, {Solitro},
  {Souami}, {Souchay}, {Spagna}, {Spina}, {Spoto}, {Steele},
  {Steidelm{\"u}ller}, {Stephenson}, {S{\"u}veges}, {Surdej}, {Szabados},
  {Szegedi-Elek}, {Taris}, {Taylor}, {Teixeira}, {Tolomei}, {Tonello}, {Torra},
  {Torra}, {Torralba Elipe}, {Trabucchi}, {Tsounis}, {Turon}, {Ulla}, {Unger},
  {Vaillant}, {van Dillen}, {van Reeven}, {Vanel}, {Vecchiato}, {Viala},
  {Vicente}, {Voutsinas}, {Weiler}, {Wevers}, {Wyrzykowski}, {Yoldas}, {Yvard},
  {Zhao}, {Zorec}, {Zucker}, \& {Zwitter}}]{2023A&A...674A...1G}
{Gaia Collaboration}, {Vallenari}, A., {Brown}, A.~G.~A., {et~al.} 2023, \aap,
  674, A1

\bibitem[{{Gokus} {et~al.}(2020){Gokus}, {Rau}, {Wilms}, {Ducci}, {Koenig},
  {Weber}, {Boller}, \& {Malyali}}]{2020ATel13657....1G}
{Gokus}, A., {Rau}, A., {Wilms}, J., {et~al.} 2020, The Astronomer's Telegram,
  13657, 1

\bibitem[{{HI4PI Collaboration} {et~al.}(2016){HI4PI Collaboration}, {Ben
  Bekhti}, {Fl{\"o}er}, {Keller}, {Kerp}, {Lenz}, {Winkel}, {Bailin},
  {Calabretta}, {Dedes}, {Ford}, {Gibson}, {Haud}, {Janowiecki}, {Kalberla},
  {Lockman}, {McClure-Griffiths}, {Murphy}, {Nakanishi}, {Pisano}, \&
  {Staveley-Smith}}]{2016A&A...594A.116H}
{HI4PI Collaboration}, {Ben Bekhti}, N., {Fl{\"o}er}, L., {et~al.} 2016, \aap,
  594, A116

\bibitem[{{Kong}(2020)}]{2020ATel13669....1K}
{Kong}, A. K.~H. 2020, The Astronomer's Telegram, 13669, 1

\bibitem[{{Kuijken} {et~al.}(2019){Kuijken}, {Heymans}, {Dvornik},
  {Hildebrandt}, {de Jong}, {Wright}, {Erben}, {Bilicki}, {Giblin}, {Shan},
  {Getman}, {Grado}, {Hoekstra}, {Miller}, {Napolitano}, {Paolilo}, {Radovich},
  {Schneider}, {Sutherland}, {Tewes}, {Tortora}, {Valentijn}, \& {Verdoes
  Kleijn}}]{2019A&A...625A...2K}
{Kuijken}, K., {Heymans}, C., {Dvornik}, A., {et~al.} 2019, \aap, 625, A2

\bibitem[{{Maccacaro} {et~al.}(1988){Maccacaro}, {Gioia}, {Wolter}, {Zamorani},
  \& {Stocke}}]{1988ApJ...326..680M}
{Maccacaro}, T., {Gioia}, I.~M., {Wolter}, A., {Zamorani}, G., \& {Stocke},
  J.~T. 1988, \apj, 326, 680

\bibitem[{{Martin} {et~al.}(2004){Martin}, {Ibata}, {Bellazzini}, {Irwin},
  {Lewis}, \& {Dehnen}}]{2004MNRAS.348...12M}
{Martin}, N.~F., {Ibata}, R.~A., {Bellazzini}, M., {et~al.} 2004, \mnras, 348,
  12

\bibitem[{{McConnachie}(2012)}]{2012AJ....144....4M}
{McConnachie}, A.~W. 2012, \aj, 144, 4

\bibitem[{{McLean} {et~al.}(2000){McLean}, {Greene}, {Lattanzi}, \&
  {Pirenne}}]{2000ASPC..216..145M}
{McLean}, B.~J., {Greene}, G.~R., {Lattanzi}, M.~G., \& {Pirenne}, B. 2000, in
  Astronomical Society of the Pacific Conference Series, Vol. 216, Astronomical
  Data Analysis Software and Systems IX, ed. N.~{Manset}, C.~{Veillet}, \&
  D.~{Crabtree}, 145

\bibitem[{{Merloni} {et~al.}(2024){Merloni}, {Lamer}, {Liu}, {Ramos-Ceja},
  {Brunner}, {Bulbul}, {Dennerl}, {Doroshenko}, {Freyberg}, {Friedrich},
  {Gatuzz}, {Georgakakis}, {Haberl}, {Igo}, {Kreykenbohm}, {Liu}, {Maitra},
  {Malyali}, {Mayer}, {Nandra}, {Predehl}, {Robrade}, {Salvato}, {Sanders},
  {Stewart}, {Tub{\'\i}n-Arenas}, {Weber}, {Wilms}, {Arcodia}, {Artis},
  {Aschersleben}, {Avakyan}, {Aydar}, {Bahar}, {Balzer}, {Becker}, {Berger},
  {Boller}, {Bornemann}, {Br{\"u}ggen}, {Brusa}, {Buchner}, {Burwitz},
  {Camilloni}, {Clerc}, {Comparat}, {Coutinho}, {Czesla}, {Dannhauer},
  {Dauner}, {Dauser}, {Dietl}, {Dolag}, {Dwelly}, {Egg}, {Ehl}, {Freund},
  {Friedrich}, {Gaida}, {Garrel}, {Ghirardini}, {Gokus}, {Gr{\"u}nwald},
  {Grandis}, {Grotova}, {Gruen}, {Gueguen}, {H{\"a}mmerich}, {Hamaus},
  {Hasinger}, {Haubner}, {Homan}, {Ider Chitham}, {Joseph}, {Joyce},
  {K{\"o}nig}, {Kaltenbrunner}, {Khokhriakova}, {Kink}, {Kirsch}, {Kluge},
  {Knies}, {Krippendorf}, {Krumpe}, {Kurpas}, {Li}, {Liu}, {Locatelli},
  {Lorenz}, {M{\"u}ller}, {Magaudda}, {Mannes}, {McCall}, {Meidinger},
  {Michailidis}, {Migkas}, {Mu{\~n}oz-Giraldo}, {Musiimenta}, {Nguyen-Dang},
  {Ni}, {Olechowska}, {Ota}, {Pacaud}, {Pasini}, {Perinati}, {Pires},
  {Pommranz}, {Ponti}, {Poppenhaeger}, {P{\"u}hlhofer}, {Rau}, {Reh},
  {Reiprich}, {Roster}, {Saeedi}, {Santangelo}, {Sasaki}, {Schmitt},
  {Schneider}, {Schrabback}, {Schuster}, {Schwope}, {Seppi}, {Serim},
  {Shreeram}, {Sokolova-Lapa}, {Starck}, {Stelzer}, {Stierhof}, {Suleimanov},
  {Tenzer}, {Traulsen}, {Tr{\"u}mper}, {Tsuge}, {Urrutia}, {Veronica},
  {Waddell}, {Willer}, {Wolf}, {Yeung}, {Zainab}, {Zangrandi}, {Zhang},
  {Zhang}, \& {Zheng}}]{2024A&A...682A..34M}
{Merloni}, A., {Lamer}, G., {Liu}, T., {et~al.} 2024, \aap, 682, A34

\bibitem[{{Momany} {et~al.}(2006){Momany}, {Zaggia}, {Gilmore}, {Piotto},
  {Carraro}, {Bedin}, \& {de Angeli}}]{2006A&A...451..515M}
{Momany}, Y., {Zaggia}, S., {Gilmore}, G., {et~al.} 2006, \aap, 451, 515

\bibitem[{{Mukai}(2017)}]{2017PASP..129f2001M}
{Mukai}, K. 2017, \pasp, 129, 062001

\bibitem[{{Predehl} {et~al.}(2021){Predehl}, {Andritschke}, {Arefiev},
  {Babyshkin}, {Batanov}, {Becker}, {B{\"o}hringer}, {Bogomolov}, {Boller},
  {Borm}, {Bornemann}, {Br{\"a}uninger}, {Br{\"u}ggen}, {Brunner}, {Brusa},
  {Bulbul}, {Buntov}, {Burwitz}, {Burkert}, {Clerc}, {Churazov}, {Coutinho},
  {Dauser}, {Dennerl}, {Doroshenko}, {Eder}, {Emberger}, {Eraerds},
  {Finoguenov}, {Freyberg}, {Friedrich}, {Friedrich}, {F{\"u}rmetz},
  {Georgakakis}, {Gilfanov}, {Granato}, {Grossberger}, {Gueguen}, {Gureev},
  {Haberl}, {H{\"a}lker}, {Hartner}, {Hasinger}, {Huber}, {Ji}, {Kienlin},
  {Kink}, {Korotkov}, {Kreykenbohm}, {Lamer}, {Lomakin}, {Lapshov}, {Liu},
  {Maitra}, {Meidinger}, {Menz}, {Merloni}, {Mernik}, {Mican}, {Mohr},
  {M{\"u}ller}, {Nandra}, {Nazarov}, {Pacaud}, {Pavlinsky}, {Perinati},
  {Pfeffermann}, {Pietschner}, {Ramos-Ceja}, {Rau}, {Reiffers}, {Reiprich},
  {Robrade}, {Salvato}, {Sanders}, {Santangelo}, {Sasaki}, {Scheuerle},
  {Schmid}, {Schmitt}, {Schwope}, {Shirshakov}, {Steinmetz}, {Stewart},
  {Str{\"u}der}, {Sunyaev}, {Tenzer}, {Tiedemann}, {Tr{\"u}mper}, {Voron},
  {Weber}, {Wilms}, \& {Yaroshenko}}]{2021A&A...647A...1P}
{Predehl}, P., {Andritschke}, R., {Arefiev}, V., {et~al.} 2021, \aap, 647, A1

\bibitem[{{Saeedi} {et~al.}(2022){Saeedi}, {Liu}, {Knies}, {Sasaki}, {Becker},
  {Bulbul}, {Dennerl}, {Freyberg}, {Laktionov}, \&
  {Merloni}}]{2022A&A...661A..35S}
{Saeedi}, S., {Liu}, T., {Knies}, J., {et~al.} 2022, \aap, 661, A35

\bibitem[{{Saeedi} \& {Sasaki}(2020)}]{2020MNRAS.499.3111S}
{Saeedi}, S. \& {Sasaki}, M. 2020, \mnras, 499, 3111

\bibitem[{{Saeedi} \& {Sasaki}(2022)}]{2022MNRAS.512.5481S}
{Saeedi}, S. \& {Sasaki}, M. 2022, \mnras, 512, 5481

\bibitem[{{Saeedi} {et~al.}(2018){Saeedi}, {Sasaki}, \&
  {Ducci}}]{2018MNRAS.473..440S}
{Saeedi}, S., {Sasaki}, M., \& {Ducci}, L. 2018, \mnras, 473, 440

\bibitem[{{Saeedi} {et~al.}(2019){Saeedi}, {Sasaki}, {Stelzer}, \&
  {Ducci}}]{2019A&A...627A.128S}
{Saeedi}, S., {Sasaki}, M., {Stelzer}, B., \& {Ducci}, L. 2019, \aap, 627, A128

\bibitem[{{Salvato} {et~al.}(2018){Salvato}, {Buchner}, {Budav{\'a}ri},
  {Dwelly}, {Merloni}, {Brusa}, {Rau}, {Fotopoulou}, \&
  {Nandra}}]{2018MNRAS.473.4937S}
{Salvato}, M., {Buchner}, J., {Budav{\'a}ri}, T., {et~al.} 2018, \mnras, 473,
  4937

\bibitem[{{Salvato} {et~al.}(2022){Salvato}, {Wolf}, {Dwelly}, {Georgakakis},
  {Brusa}, {Merloni}, {Liu}, {Toba}, {Nandra}, {Lamer}, {Buchner}, {Schneider},
  {Freund}, {Rau}, {Schwope}, {Nishizawa}, {Klein}, {Arcodia}, {Comparat},
  {Musiimenta}, {Nagao}, {Brunner}, {Malyali}, {Finoguenov}, {Anderson},
  {Shen}, {Ibarra-Medel}, {Trump}, {Brandt}, {Urry}, {Rivera}, {Krumpe},
  {Urrutia}, {Miyaji}, {Ichikawa}, {Schneider}, {Fresco}, {Boller}, {Haase},
  {Brownstein}, {Lane}, {Bizyaev}, \& {Nitschelm}}]{2022A&A...661A...3S}
{Salvato}, M., {Wolf}, J., {Dwelly}, T., {et~al.} 2022, \aap, 661, A3

\bibitem[{{Sazonov} {et~al.}(2006){Sazonov}, {Revnivtsev}, {Gilfanov},
  {Churazov}, \& {Sunyaev}}]{2006A&A...450..117S}
{Sazonov}, S., {Revnivtsev}, M., {Gilfanov}, M., {Churazov}, E., \& {Sunyaev},
  R. 2006, \aap, 450, 117

\bibitem[{{Schlafly} \& {Finkbeiner}(2011)}]{2011ApJ...737..103S}
{Schlafly}, E.~F. \& {Finkbeiner}, D.~P. 2011, \apj, 737, 103

\bibitem[{{Secrest} {et~al.}(2015){Secrest}, {Dudik}, {Dorland}, {Zacharias},
  {Makarov}, {Fey}, {Frouard}, \& {Finch}}]{2015ApJS..221...12S}
{Secrest}, N.~J., {Dudik}, R.~P., {Dorland}, B.~N., {et~al.} 2015, \apjs, 221,
  12

\bibitem[{{Smith} {et~al.}(2001){Smith}, {Brickhouse}, {Liedahl}, \&
  {Raymond}}]{2001ApJ...556L..91S}
{Smith}, R.~K., {Brickhouse}, N.~S., {Liedahl}, D.~A., \& {Raymond}, J.~C.
  2001, \apjl, 556, L91

\bibitem[{{Stiele} {et~al.}(2011){Stiele}, {Pietsch}, {Haberl},
  {Hatzidimitriou}, {Barnard}, {Williams}, {Kong}, \&
  {Kolb}}]{2011A&A...534A..55S}
{Stiele}, H., {Pietsch}, W., {Haberl}, F., {et~al.} 2011, \aap, 534, A55

\bibitem[{{Tolstoy} {et~al.}(2009){Tolstoy}, {Hill}, \&
  {Tosi}}]{2009ARA&A..47..371T}
{Tolstoy}, E., {Hill}, V., \& {Tosi}, M. 2009, \araa, 47, 371

\bibitem[{{van den Eijnden} {et~al.}(2020){van den Eijnden}, {Degenaar},
  {Wijnands}, {Russell}, {Stoop}, {Armas Padilla}, {Russell}, {Maitra},
  {Heinke}, {Sivakoff}, {Shaw}, {Maccarone}, {Miller-Jones}, \&
  {Bahramian}}]{2020ATel13716....1V}
{van den Eijnden}, J., {Degenaar}, N., {Wijnands}, R., {et~al.} 2020, The
  Astronomer's Telegram, 13716, 1

\bibitem[{{Wenger} {et~al.}(2000){Wenger}, {Ochsenbein}, {Egret}, {Dubois},
  {Bonnarel}, {Borde}, {Genova}, {Jasniewicz}, {Lalo{\"e}}, {Lesteven}, \&
  {Monier}}]{2000A&AS..143....9W}
{Wenger}, M., {Ochsenbein}, F., {Egret}, D., {et~al.} 2000, \aaps, 143, 9

\bibitem[{{Willingale} {et~al.}(2013){Willingale}, {Starling}, {Beardmore},
  {Tanvir}, \& {O'Brien}}]{2013MNRAS.431..394W}
{Willingale}, R., {Starling}, R.~L.~C., {Beardmore}, A.~P., {Tanvir}, N.~R., \&
  {O'Brien}, P.~T. 2013, \mnras, 431, 394

\bibitem[{{Wright} {et~al.}(2010){Wright}, {Eisenhardt}, {Mainzer}, {Ressler},
  {Cutri}, {Jarrett}, {Kirkpatrick}, {Padgett}, {McMillan}, {Skrutskie},
  {Stanford}, {Cohen}, {Walker}, {Mather}, {Leisawitz}, {Gautier}, {McLean},
  {Benford}, {Lonsdale}, {Blain}, {Mendez}, {Irace}, {Duval}, {Liu}, {Royer},
  {Heinrichsen}, {Howard}, {Shannon}, {Kendall}, {Walsh}, {Larsen}, {Cardon},
  {Schick}, {Schwalm}, {Abid}, {Fabinsky}, {Naes}, \&
  {Tsai}}]{2010AJ....140.1868W}
{Wright}, E.~L., {Eisenhardt}, P. R.~M., {Mainzer}, A.~K., {et~al.} 2010, \aj,
  140, 1868

\bibitem[{{Yang} {et~al.}(2022){Yang}, {Hare}, {Kargaltsev}, {Volkov}, {Chen},
  \& {Rangelov}}]{2022ApJ...941..104Y}
{Yang}, H., {Hare}, J., {Kargaltsev}, O., {et~al.} 2022, \apj, 941, 104

\end{thebibliography}

\begin{appendix}
\onecolumn
\begin{landscape}

\section{Catalouges}
\raggedright
\label{final-cata}
\setlength{\tabcolsep}{1.1mm}
{\scriptsize
\begin{table*}
\centering
    \caption{Catalouge of X-ray sources in the FOV of CMa\,OD (extract).$^{\dagger}$  \label{catalogue-x-ray}}
\begin{tabular}{cccccccccccll}
\hline
No & \erosita\, SRC-Name &RA              &DEC              &r1$\sigma$                       & Flux\,(0.2--5.~keV) &ML&HR1&HR2&HR3&HR4&Class$^{*}$&Comment\\
                       &   &  (J2000)       &  (J2000)        & ($\arcsec$)                     & $10^{-14}$ erg\,s$^{-1}$\,cm$^{-2}$)&&&&&&&\\        
\hline
1&    1eRASS J063832.0-164540& 06 38 32.03& -16 45 40.8& 4.00& 8.94E-14$\pm$3.09E-14& 20.82&--&--&--&--& AGN&\\
2&    1eRASS J063838.3-164956& 06 38 38.39& -16 49 56.5& 3.25& 1.41E-13$\pm$3.90E-14& 38.36   &--&0.09$\pm$0.28&--&--&AGN&\\
3&    1eRASS J063839.6-164412& 06 38 39.63& -16 44 12.9& 4.91& 3.66E-14$\pm$1.91E-14& 7.74    &--&--&--&--&FG& SIMBAD: TYC 5948-2725-1\\
4&    1eRASS J063847.8-172310& 06 38 47.82& -17 23 10.1& 5.10& 5.32E-14$\pm$2.41E-14& 10.54   &--&--&--&--&AGN&\\
5&    1eRASS J063851.0-161150& 06 38 51.00& -16 11 50.7& 5.05& 3.06E-14$\pm$1.78E-14& 6.30    &--&--&--&--&-- &\\
6&    1eRASS J063858.1-164847& 06 38 58.10& -16 48 47.4& 4.53& 3.59E-14$\pm$1.95E-14& 6.78    &--&--&--&--&--&\\
7&    1eRASS J063859.7-174517& 06 38 59.79& -17 45 17.5& 4.81& 4.92E-14$\pm$2.31E-14& 10.90   &--&--&--&--&FG& SIMBAD: ATO J099.7479-17.7554\\
8&    1eRASS J063907.5-155410& 06 39 07.57& -15 54 10.2& 3.86& 7.76E-14$\pm$2.96E-14& 19.20   &--&0.05$\pm$0.41&--&--&AGN& SIMBAD: PMN J0639-1554\\
9&    1eRASS J063908.2-155555& 06 39 08.22& -15 55 55.4& 5.67& 3.90E-14$\pm$2.09E-14& 6.48    &--&--&--&--&--&\\
10&   1eRASS J063908.5-165212& 06 39 08.58& -16 52 12.4& 4.18& 4.07E-14$\pm$2.08E-14& 9.28    &--&--&--&--&STRL&\\
11&   1eRASS J063920.0-155101& 06 39 20.07& -15 51 01.3& 4.84& 5.30E-14$\pm$2.49E-14& 9.23    &--&--&--&--&FG& SIMBAD: TYC 5948-2389-1\\
12&   1eRASS J063924.7-163408& 06 39 24.78& -16 34 08.8& 5.28& 4.17E-14$\pm$2.24E-14& 7.03    &--&--&--&--&--&\\
13&   1eRASS J063925.0-163635& 06 39 25.06& -16 36 35.6& 7.59& 3.82E-14$\pm$2.09E-14& 6.18    &--&--&--&--& AGN&\\
14&   1eRASS J063925.1-164145& 06 39 25.14& -16 41 45.2& 5.73& 5.98E-14$\pm$3.00E-14& 10.81   &--&--&--&--&AGN&\\
15&   1eRASS J063928.1-165406& 06 39 28.10& -16 54 06.5& 7.99& 4.14E-14$\pm$2.22E-14& 6.16    &--&--&--&--&AGN&\\
16&   1eRASS J063935.4-155944& 06 39 35.50& -15 59 44.6& 3.66& 5.42E-14$\pm$2.41E-14& 14.53   &--&--&--&--&FG& SIMBAD: ASAS J063936-1559.8\\
17&   1eRASS J063936.8-180249& 06 39 36.89& -18 02 49.8& 4.53& 3.52E-14$\pm$2.04E-14& 6.28    &--&--&--&--&--&\\
18&   1eRASS J063937.2-163926& 06 39 37.23& -16 39 26.1& 5.04& 3.65E-14$\pm$2.13E-14& 6.39    &--&--&--&--&FG&SIMBAD: TYC 5948-1925-1\\
19&   1eRASS J063947.3-175218& 06 39 47.33& -17 52 18.8& 5.28& 5.17E-14$\pm$2.40E-14& 11.63   &--&0.12$\pm$0.43&--&--&--&\\
20&   1eRASS J063949.4-154343& 06 39 49.44& -15 43 43.4& 4.82& 3.24E-14$\pm$1.89E-14& 6.04    &--&--&--&--&--&\\
...&   & & & & & &&&&&&\\
\hline
      \hline
      ~\\
    \multicolumn{13}{l}{$\dagger$:\footnotesize	The full catalogue is available at the CDS. The table presents the ID of source in this work (NO), \erosita\, source name in {\citet{2024A&A...682A..34M}} catalogue, Right ascension\,(RA), }\\
    \multicolumn{13}{l}{\footnotesize ~~~Declination\,(DEC), 1$\sigma$ positional error of the source\,(r), source flux, Detection maximum likelihood\,(ML), significant hardness ratios of sources based on the energy bands which are }\\
    \multicolumn{13}{l}{\footnotesize ~~~defined in Sect.\,\ref{hr-sec}, classification of the source based on this work, and comments column presents the SIMBAD ID of the sources as exist in SIMBAD data base, plus the information is }\\
    \multicolumn{13}{l}{\footnotesize ~~~source has the problem of optical loading. }\\
    \multicolumn{13}{l}{\footnotesize *: AGN:\,Active galactic nuclei; FG:\,Foreground star; SYM:\,Symbiotic star; STRL:\,stellar object; Cma\,OD:\,candidate Member of Cma\,OD; Can:\,Candidate }\\
\hline
\end{tabular}    
\end{table*}}
\end{landscape}

\pagebreak
\label{counterparts-cata}
\setlength{\tabcolsep}{1.1mm}
{\small
\begin{table*}
\centering
    \caption{ WISE and \gaia\, counterpart of X-ray sources in the field of Cma\,OD.$^{ \dagger}$  \label{catalogue-x-ray}}
\begin{tabular}{cccccccc}
\hline
No & W1 &  W2            &W3             & W4& G-mag& RG-0mag& BG-mag\\         
\hline
1&14.07$\pm$0.02& 12.85 0.02& 9.94 $\pm$ 0.05& 7.30 $\pm$0.11& 18.81$\pm$0.01& 18.28 $\pm$0.03& 19.09 $\pm$0.05\\
2& 15.74 $\pm$0.05& 14.80 $\pm$0.06&  12.00 $\pm$ 0.29&   8.28 &   20.39$\pm$   0.02&  19.64$\pm$ 0.12&20.84$\pm$ 0.24\\
3& 10.13 $\pm$0.02& 10.16 $\pm$0.02&10.11$\pm$ 0.06& 8.95$\pm$ 0.46 &11.94$\pm$ 0.01& 11.36$\pm$ 0.01& 12.36 $\pm$0.01\\
4&15.53 $\pm$0.04& 14.91$\pm$0.07& 12.08$\pm$0.34& 8.96$\pm$ 0.51 &21.01$\pm$ 0.02& 19.88 $\pm$0.24& 20.64$\pm$ 0.13\\
5& 15.24  $\pm$ 0.04  &  15.13 $\pm$  0.07&     12.10     &        8.88 & 17.76$\pm$      0.01& 17.09$\pm$        0.02& 18.18$\pm$        0.02\\
6& --&--&--&--&--&--&--\\
7 &8.29 $\pm$0.02& 8.33 $\pm$0.02& 8.23 $\pm$0.02 &8.34 $\pm$0.27 &11.35 $\pm$0.01&10.49 $\pm$0.01& 12.08$\pm$0.01\\
8& 14.49$\pm$0.02& 13.32$\pm$0.03& 10.56 $\pm$0.08& 8.11 $\pm$0.21 &  --& -- &--\\
9& 12.35 $\pm$  0.02&   12.39$\pm$      0.02&   12.36  &        9.02 &14.76 $\pm$0.01 &14.04$\pm$0.01 &15.32 $\pm$0.01\\
10& 11.73$\pm$0.02&     11.71$\pm$ 0.02&        11.28$\pm$      0.15&   8.60 & 14.34 $\pm$0.01 &13.43$\pm$0.01& 14.76$\pm$0.01\\
11& 10.42$\pm$0.02& 10.42 $\pm$0.02& 10.29 $\pm$0.06& 8.84& 11.83 $\pm$0.01& 11.35 $\pm$0.01& 12.14$\pm$0.01\\
12& 14.66 $\pm$ 0.03&   14.53$\pm$      0.05&   12.71& 9.02&19.08$\pm$   0.01 &17.70 $\pm$       0.01 &20.59 $\pm$       0.13\\
13& 14.62$\pm$0.03& 14.33$\pm$0.05 &12.23 $\pm$0.34 &8.86  &12.49$\pm$0.01&11.48 $\pm$0.01& 13.31$\pm$0.01\\
14& -- & --& --& -- &20.48$\pm$0.02& 9.84$\pm$0.14& 20.80$\pm$0.20\\
15& 13.73 $\pm$0.033& 12.61 $\pm$0.02& 9.95 $\pm$0.05& 7.70 $\pm$0.14& 18.92$\pm$0.01&  19.16$\pm$0.10& 18.17$\pm$0.05\\
16 &10.78$\pm$0.02& 10.80 $\pm$0.02& 10.80$\pm$0.11 &8.88&13.10 $\pm$0.01& 12.37$\pm$0.03 &13.54 $\pm$0.04\\
17& 16.79 $\pm$ 0.10&   17.01$\pm$      0.35&12.486&    8.822 &-- &-- &-- \\
18& 8.64 $\pm$0.02& 8.66 $\pm$0.02 &8.60 $\pm$0.02 &8.12$\pm$0.21 &9.70 $\pm$0.01& 9.32 $\pm$0.01 &9.93$\pm$0.01\\
19& 12.04$\pm$0.02& 12.07 $\pm$0.021& 12.57 $\pm$0.43& 8.88& 14.30 $\pm$0.01& 13.62 $\pm$0.01& 14.83 $\pm$0.01\\
20& 11.55$\pm$0.02& 11.56 $\pm$0.02 &11.49 $\pm$0.18& 9.12& 18.95 $\pm$0.01& 18.20 $\pm$0.02& 19.58 $\pm$0.03\\

...&   & & & & &&\\
      \hline
      ~\\
    \multicolumn{8}{l}{$\dagger$: Full catalogue is available at the CDS. The table shows the magnitudes of WISE and \gaia\,counterpart of the X-ray sources}\\
    \multicolumn{8}{l}{~~~(see Sects.\,\ref{infra-sec} and\,\ref{optical-sec}) The magnitudes without errors are only upper limits.} \\    
\hline
\end{tabular}
\end{table*}}

\end{appendix}


\end{document}